\documentclass[11pt,a4paper]{article}
\usepackage[T1]{fontenc}
\usepackage{lmodern}
\usepackage[margin=25mm]{geometry}
\usepackage{amsmath,amssymb,amsthm,mathtools,mathrsfs}
\usepackage{microtype,enumitem,booktabs,needspace}
\usepackage{graphicx,tabularx,array}
\graphicspath{{figures/}}
\usepackage[hidelinks]{hyperref}
\allowdisplaybreaks[2]
\setlist{nosep,leftmargin=1.8em}
\newtheorem{assumption}{Assumption}
\newtheorem{theorem}{Theorem}
\newtheorem{corollary}{Corollary}
\newtheorem{lemma}{Lemma}
\newtheorem{proposition}{Proposition}
\theoremstyle{definition}
\newtheorem{remark}{Remark}

\newcommand{\R}{\mathbb R}
\newcommand{\E}{\mathbb E}
\newcommand{\PP}{\mathbb P}
\newcommand{\F}{\mathscr F}
\newcommand{\Et}{\mathbb E_t}
\newcommand{\Reg}{\mathcal R}
\newcommand{\epsopt}{\varepsilon^*}

\title{Adaptive Control for General Nonlinear Stochastic Systems}
\author{Haizheng Li\thanks{This work was supported by the National Natural Science Foundation of China under Grant No. 12288201.}
		\thanks{Haizheng Li is with the State Key Laboratory of Mathematical Sciences, Academy of Mathematics and Systems Science, Chinese Academy of Sciences, Beijing 100190, China, and also with the School of Mathematical Science, University of Chinese Academy of Sciences, Beijing 100049, China. (e-mail: lihaizheng@amss.ac.cn). }}
\date{}

\begin{document}
\maketitle
\begin{abstract}
This paper extends the least-squares and certainty-equivalence
minimum-variance framework of classical linear self-tuning regulators
to a broad class of nonlinearly parameterized systems with compact
parameter spaces, including commonly used deep neural network models.
First, historical nonlinear least-squares fitting guarantees
are converted into online one-step prediction bounds under a general
modulus of continuity, with explicit regret rates under parameter
Lipschitz continuity and finite-moment noise.
Second,
approximate NLS estimation combined with certainty-equivalence control
achieves closed-loop stability and asymptotically optimal tracking under mild conditions without requiring excitation or parameter identifiability.
Explicit closed-loop regret and tracking error rates are established
in the parameter-Lipschitz case, and closed-loop examples show that
the regret rates cannot, in general, be improved for this regulator.
Finally, robustness to small unmodeled dynamics is established by showing that closed-loop stability is preserved and the tracking error bounds comprise a vanishing term and a residual term determined by the model discrepancy.
\end{abstract}


\section{Introduction}\label{sec:introduction}

\subsection{Background and motivation}

Modern applications in embodied robotics, intelligent manufacturing, and energy systems increasingly combine feedback control with dynamical models that are estimated from data and may be highly nonlinear \cite{seo2024avoiding}. Accurate models may be difficult to obtain before operation, while measured input--output data can be used to update the model and improve control performance during operation. The self-tuning regulator (STR) provides a direct way to incorporate this information into feedback: estimate the model from past observations and compute the next control input using the current estimate. Its classical realization combines least-squares (LS) estimation with certainty-equivalence (CE)  control \cite{ASTROM1973,chenguo1991}. Extending this approach to nonlinear model classes would provide a theoretical basis for adaptive regulation using general nonlinear models estimated from data.

The central difficulty is the interaction between open-loop prediction and closed-loop feedback control. The estimated model determines the control input, which in turn affects the subsequent system trajectory and the observations available for estimation. The predictor must therefore remain accurate along the trajectory generated by the adaptive regulator. For nonlinear parameterizations, a small fitting error over past observations alone does not ensure a small one-step prediction error, and feedback may amplify prediction errors before sufficient information has been obtained. Stability and tracking analysis must therefore account for prediction errors before boundedness of the closed-loop signals has been established.

This paper studies general nonlinearly parameterized stochastic systems with finite output and input memory, where the unknown parameter belongs to a known compact set. At each time, the estimator approximately minimizes the cumulative squared output residuals, and the control input is determined according to the CE principle. The objective is to establish closed-loop stability and asymptotically optimal tracking for this nonlinear STR. To this end, a one-step prediction analysis is developed from the historical least-squares comparison, allowing the regressors to be unbounded, dependent, and nonstationary.

\subsection{Related work}

For linear stochastic systems, the stability and optimality of the
\AA{}str\"om--Wittenmark self-tuning regulator and an extended
least-squares adaptive tracker were established in~\cite{guo1991}.
A unified treatment of stability, optimality, consistency, and
logarithmic convergence rates for LS-based regulators was subsequently
developed in~\cite{guo1995}. These classical works provide a
methodological foundation for the present study through their analysis
of the interaction between prediction errors and closed-loop signal
growth. 

However, extensions to nonlinear systems are subject to fundamental limitations that apply to any causal controller. A critical growth phenomenon for scalar-parameter nonlinear stochastic systems was identified in~\cite{guo1997}, while limitations arising from uncertainty growth and parameter dimension were established in~\cite{xieguo1999,xieguo2000uncertainty}. For nonlinear parameterizations, feedback capability was related to parameter sensitivity and model structure in~\cite{liguo2011}. These results motivate the search for conditions under which nonlinear adaptive regulation can be achieved. 

Early work on nonlinear adaptive prediction
in~\cite{laizhu1991adaptive} studied nonlinear autoregressive and
control models with unknown parameters in a compact set and
established an asymptotically exact logarithmic law for cumulative nonlinear least squares (NLS) one-step prediction errors under smoothness,
identification, and a positive-definite limiting information
condition. Subsequently, \cite{lichen2016} used NLS to achieve
simultaneous identification and stabilization for scalar-parameter
systems. In the analysis of control regret, \cite{boffi2021}
related prediction regret to control regret for linearly
parameterized systems under suitable incremental stability
conditions. More recently, \cite{liuli2025} established asymptotically
optimal output regulation with stable, nonlinearly parameterized
exosystems and derived NLS identification rates, while
\cite{wang2025logarithmic} obtained polylogarithmic expected regret
across episodes for nonlinearly parameterized systems under a
uniformly bounded dynamics map, an identifying initial policy,
and excitation of the optimal policy.
Recent recursive approaches include a nonlinear WLS-type estimator
and adaptive feedback under quantitative monotonicity and Lipschitz
conditions~\cite{zhang2026stochastic}, and a modified
stochastic-gradient predictor with prediction and CE control rates
under a weak-convexity condition on the prediction
loss~\cite{liu2026gradient}. The present work establishes stability and tracking properties of an adaptive regulator that combines approximate cumulative NLS estimation with CE control for general nonlinearly parameterized systems. The analysis requires neither excitation nor parameter convergence and does not impose additional model structure conditions such as uniformly bounded dynamics, identifying information conditions, or monotonicity or convexity assumptions used in the respective studies above.

From a statistical learning perspective,
\cite{stamouli2024pem} obtained rate-optimal non-asymptotic bounds
for the quadratic prediction error method under dependence and
identifiability conditions. Their performance criterion evaluates the final model estimate on an independent copy of the regressor trajectory. The criterion considered here is the cumulative one-step prediction error along the actual system trajectory, with the model estimate updated at each time. Under heavy-tailed errors,
\cite{hanwellner2019,kuchibhotlapatra2022} characterized how noise
moments, function-class complexity, and local envelopes govern
least-squares rates. Those results concern statistical prediction risk based on independent sample pairs. The present analysis allows predictable regressors generated from past observations, as occurs under feedback control. 

\subsection{Main contributions and organization}

The main contributions are as follows.
\begin{enumerate}[label=\textup{(\roman*)}]
\item
One-step prediction error bounds are established for predictable nonlinear model families with compact parameter sets. Approximate least-squares estimation based on past observations yields cumulative prediction error bounds under a general modulus of continuity. Under parameter Lipschitz continuity, explicit almost-sure convergence rates are obtained for finite-moment noise. The bounds account for the accumulated energy, peak values, and higher-order sample moments of the possibly unbounded prediction envelope.

\item
With the CE control input, the prediction bounds establish time-average closed-loop stability and almost-sure convergence of the average squared tracking error to the noise variance, the asymptotically minimum achievable level. Explicit convergence rates are obtained under parameter Lipschitz continuity. Closed-loop examples show that the derived regret convergence rates cannot, in general, be improved for the regulator considered.

\item
The robustness of the same adaptive regulator to small unmodeled dynamics is established. Closed-loop stability is preserved when the model discrepancy is sufficiently small. The tracking error bounds contain a vanishing term and a residual term due to the unmodeled dynamics.
\end{enumerate}

Section~\ref{sec:formulation} presents the system assumptions and
performance criteria. Section~\ref{sec:algorithm} specifies the
estimator and controller. The main results and their proofs are
given in Sections~\ref{sec:results} and~\ref{sec:proofs}, respectively.
Section~\ref{sec:conclusion} concludes the paper. Proofs of the lower bounds on the closed-loop regret and of the auxiliary martingale lemmas are given in Appendices~\ref{app:sharpness} and~\ref{app:offset-proof}, respectively.

\section{Problem Formulation}\label{sec:formulation}
\subsection{Notation}
The symbols $\R$, $\E$, and $\PP$ denote the real numbers, expectation,
and probability, respectively. For a filtration $(\F_t)$, write
$\Et[\cdot]=\E[\cdot\mid\F_t]$. A quantity selected at time $t$ is
called predictable here when it is $\F_t$-measurable and is used to
predict the observation at time $t+1$. Write $x_+=\max\{x,0\}$,
$a\vee b=\max\{a,b\}$, and $a\wedge b=\min\{a,b\}$. The notation $a_T=O(b_T)$ almost surely means that, on an event of
probability one, $|a_T|\le Cb_T$ eventually for some finite
path-dependent constant $C$; $a_T=o(b_T)$ means $a_T/b_T\to0$ on
that event. The relation $a_T\asymp b_T$ means that their
ratio is bounded above and below by positive constants. A positive
function $\ell$ is slowly varying if
$\ell(cx)/\ell(x)\to1$ as $x\to\infty$ for every fixed $c>0$.
Constants denoted by $C$ may change from one occurrence to the next.
\subsection{System and assumptions}

Consider the nonlinear autoregressive system with exogenous inputs (NARX) described by
\begin{equation}\label{eq:plant}
 y_{t+1}=f(\theta^*,Y_t,U_t)+w_{t+1},\qquad t\ge0,
\end{equation}
where, for fixed integers $p\ge1$ and $q\ge0$,
\begin{equation}\label{eq:regressors}
 Y_t=(y_t,\cdots,y_{t-p+1})^\top,\quad
 U_t^-=(u_{t-1},\cdots,u_{t-q})^\top,\quad
 U_t=(u_t,(U_t^-)^\top)^\top.
\end{equation}
When $q=0$, $U_t^-$ is the empty vector with norm zero.
The finite initial input and output histories are deterministic, and we set $w_0=0$
for the signal-growth estimates below.

The system is defined on a complete filtered probability space with filtration $(\F_t)$.
The initial histories and the random variables $Y_t,U_t^-,r_{t+1}$ are
$\F_t$-measurable. The parameter estimate and control input selected at time $t$ are also
$\F_t$-measurable, whereas $w_{t+1}$ is $\F_{t+1}$-measurable. Thus, the
control input is determined before observing $w_{t+1}$ or $y_{t+1}$. The
reference signal satisfies $|r_{t+1}|\le\bar r$ for a finite $\bar r$.

\begin{assumption}[Parameter continuity and linear growth of model differences]
\label{ass:family}
The known nonempty compact set $\Theta\subset\R^d$, $d\ge1$, contains
the true parameter $\theta^*$ and has diameter at most $R>0$.
The finite real-valued function
$f:\Theta\times\R^p\times\R^{q+1}\to\R$ is jointly Borel measurable.
There is a finite nondecreasing function
$\omega:[0,\infty)\to[0,\infty)$ with
\[
 \omega(0)=0,\qquad \lim_{h\to0^+}\omega(h)=0,
\]
such that, for all parameters and all regressors,
\begin{equation}\label{eq:parammod}
{\quad
 |f(\theta,Y,U)-f(\vartheta,Y,U)|
 \le (1+\|Y\|+\|U\|)\omega(\|\theta-\vartheta\|).
 \quad}
\end{equation}
\end{assumption}
\begin{remark}\label{rem:family-scope}
The linear growth bound on model differences in Assumption~1 is motivated by fundamental limitations of feedback: for any $\upsilon>1$, an envelope of the form $(1+|Y|+|U|)^\upsilon$ admits model classes of sufficiently large parameter dimension that no causal feedback can globally stabilize~\cite{xieguo1999}, precluding unrestricted
guarantees of asymptotically optimal tracking. This growth bound also accommodates multilayer feedforward neural
networks, with admissible activations including ReLU, leaky ReLU
with a fixed slope, $\tanh$, the logistic sigmoid, and softplus; see Remark \ref{rem:nn-unmodeled} for details. The continuity condition on the parameters is distinct from a Lipschitz condition
on the regressors. For nonparametric model classes, even global
Lipschitz continuity in the regressors can be insufficient for global
feedback stabilization~\cite{zhangguo2002}, which highlights the importance of the assumed model structure for stabilization.
\end{remark}
\begin{assumption}[Model inversion and the minimum-phase property]\label{ass:reach}
There exist a jointly Borel map
\(
 \psi:\Theta\times\R^p\times\R^q\times\R\longrightarrow\R
\)
and a finite constant $K$, possibly depending on $\psi$, such that, almost surely, for every
$t\ge0$ and $\theta\in\Theta$,
\begin{align}
 f\bigl(\theta,Y_t,(\psi(\theta,Y_t,U_t^-,r_{t+1}),(U_t^-)^\top)^\top\bigr)
 &=r_{t+1},\label{eq:reach}\\
 |\psi(\theta,Y_t,U_t^-,r_{t+1})|
 &\le K(1+\|Y_t\|+\|U_t^-\|).\label{eq:invgrowth}
\end{align}
Moreover, when $q\ge1$, there is a fixed constant
$\lambda\in(0,1)$ such that, along the resulting closed-loop trajectory, the following minimum-phase condition holds
\begin{equation}\label{eq:mp}
 u_{t-1}^2
 =O\!\left(1+\sum_{s=0}^{t}\lambda^{t-s}(y_s^2+w_s^2)\right)
 \qquad\text{a.s.}
\end{equation}
\end{assumption}

\begin{remark}
\label{rem:inverse-sufficient}
A sufficient condition for the existence of such $\psi(\cdot)$ is differentiability with respect to the current scalar
input together with the following bounds, for some constants $b>0$ and $C_0<\infty$,
\begin{align}
 \partial_u f\bigl(\theta,Y,(u,(U^-)^\top)^\top\bigr)
 &\ge b,\label{eq:inverse-slope}\\
 \bigl|f\bigl(\theta,Y,(0,(U^-)^\top)^\top\bigr)\bigr|
 &\le C_0(1+\|Y\|+\|U^-\|),\label{eq:inverse-offset}
\end{align}
uniformly over their domains \cite{huangguo2012}. For a specified rule for selecting parameter estimates, it is not necessary that every
parameter in $\Theta$ admit such an inverse map. It suffices that the
estimates actually selected admit a measurable inverse at the prescribed
reference, with a common linear growth bound uniform in time, as in the estimate-modification methods of, e.g., \cite{guo1991,guo1995}. The certainty-equivalence control law can also be modified to improve control performance, as in the cautious control approach of \cite{zhang2006cautious}, while retaining similar convergence guarantees. The minimum-phase condition \eqref{eq:mp} is necessary for the internal stability of the closed-loop system in the linear case, as discussed in \cite{guo1995}.
\end{remark}

\begin{assumption}[Noise]\label{ass:noise}
For some $\beta>2$ and a
finite constant $\sigma^2\ge0$,
\begin{equation}\label{eq:noise}
 \Et w_{t+1}=0,\qquad \Et w_{t+1}^2=\sigma^2,\qquad
 \sup_{t\ge0}\Et|w_{t+1}|^{\beta}<\infty\quad\text{a.s.}
\end{equation}
\end{assumption}

Write $W_T=\sum_{t<T}w_{t+1}^2$ for the noise energy.

\subsection{Prediction and tracking criteria}

Along the system trajectory, define
\begin{equation}\label{eq:DG}
 D_t(\theta)=f(\theta,Y_t,U_t)-f(\theta^*,Y_t,U_t),\qquad
 \Delta_t=D_t(\hat\theta_t).
\end{equation}
Under the control law specified in
Section~\ref{sec:algorithm}, the closed-loop dynamics satisfy
\begin{equation}\label{eq:closedloop}
 y_{t+1}=r_{t+1}-\Delta_t+w_{t+1}.
\end{equation}
Define the cumulative squared prediction error (regret) and the time-average squared tracking error by
\begin{equation}\label{eq:performance}
 \Reg_T=\sum_{t<T}\Delta_t^2,\qquad
 \mathcal J_T=\frac1T\sum_{t<T}(y_{t+1}-r_{t+1})^2,\qquad T\ge1.
\end{equation}

\section{Adaptive Control Algorithm}\label{sec:algorithm}
\subsection{Nonlinear least squares and model inversion}

Define the cumulative least-squares criterion
\begin{equation}\label{eq:Q}
 Q_t(\theta)=\sum_{s=0}^{t-1}
 [y_{s+1}-f(\theta,Y_s,U_s)]^2,\qquad Q_0\equiv0.
\end{equation}
Fix any finite optimization tolerance $\epsopt\ge0$. The adaptive regulator is specified as follows.

\noindent\textbf{Algorithm 1.}
Given $\Theta$, the initial histories, and $\epsopt$, set $Q_0\equiv0$.
At each time $t$, before observing $y_{t+1}$, perform the following steps:
\begin{enumerate}
\item Choose an $\F_t$-measurable estimate satisfying
\begin{equation}\label{eq:algorithm}
  {\quad
 Q_t(\hat\theta_t)\le\min_{\theta\in\Theta}Q_t(\theta)+\epsopt,\quad \hat\theta_t\in \Theta.
 \quad}
\end{equation}
\item Using an inverse map supplied by Assumption~\ref{ass:reach}, compute the CE control input as
\begin{equation}\label{eq:algorithmcontrol}
  {\quad u_t=\psi(\hat\theta_t,Y_t,U_t^-,r_{t+1}).\quad}
\end{equation}
\end{enumerate}
Apply $u_t$, observe $y_{t+1}$, and update
\[
 Q_{t+1}(\theta)=Q_t(\theta)+[y_{t+1}-f(\theta,Y_t,U_t)]^2.
\]

For every finite data record, the criterion is continuous on the compact set $\Theta$,
so a minimum is attained and a measurable minimizer can be selected. The specified
Borel inverse map yields a finite predictable control input, and the finite-valued
system dynamics yield finite observations at each subsequent time. Every selected estimate satisfies
\begin{equation}\label{eq:comparator}
 Q_t(\hat\theta_t)\le Q_t(\theta^*)+\epsopt=W_t+\epsopt.
\end{equation}

\begin{remark}\label{rem:computation} For any $\varepsilon^*>0$, an estimate satisfying \eqref{eq:algorithm} can be obtained by the RSAGD algorithm in \cite{wangguo2017} in finitely many iterations. In addition, if finite covers and the bounds in
Assumption~\ref{ass:family} can be evaluated, a finite search yields an estimate within
any prescribed tolerance $\epsopt>0$ at each finite time.
Indeed, fix $\theta_0\in\Theta$ and set\(
 a_s=|y_{s+1}-f(\theta_0,Y_s,U_s)|
       +(1+\|Y_s\|+\|U_s\|)\omega(R).
\)
Every residual has magnitude at most $a_s$, so
\(
 |Q_t(\theta)-Q_t(\vartheta)|
 \le 2\omega(\|\theta-\vartheta\|)
       \sum_{s<t}a_s(1+\|Y_s\|+\|U_s\|).
\)
Minimization over a $\delta$-net therefore satisfies~\eqref{eq:algorithm}
whenever
\(2\omega(\delta)\sum_{s<t}a_s(1+\|Y_s\|+\|U_s\|)\le\epsopt.
\)
Such a resolution exists at every finite time.
\end{remark}
\begin{remark}\label{parameterconstraint} The hard constraint $\hat\theta_t\in\Theta$ in \eqref{eq:algorithm} can be replaced by a
regularization term, provided that the model and inverse map are defined
on $\mathbb R^d$ and Assumption~\ref{ass:family} and the inversion conditions~\eqref{eq:reach}--\eqref{eq:invgrowth} hold uniformly on every bounded parameter set.
Specifically, initialize $\hat\theta_0\in\Theta$ and, for $t\ge1$,
choose an approximate minimizer of
\(
 \tilde{Q}_t(\theta)
 \triangleq Q_t(\theta)+\gamma t\,\operatorname{dist}(\theta,\Theta),\gamma>0,
\)
such that
\(
 \tilde{Q}_t(\hat\theta_t)
 \le \inf_{\theta\in\mathbb R^d}\tilde{Q}_t(\theta)+\varepsilon^*.
\)
Since $\theta^*\in\Theta$,
\(
 Q_t(\hat\theta_t)
 +\gamma t\,\operatorname{dist}(\hat\theta_t,\Theta)
 \le Q_t(\theta^*)+\varepsilon^*
 =\sum_{s<t}w_{s+1}^2+\varepsilon^*.
\)
Consequently,
\[
 \operatorname{dist}(\hat\theta_t,\Theta)
 \le
 \frac{\sum_{s<t}w_{s+1}^2+\varepsilon^*}{\gamma t}.
\]
The noise strong law and compactness of $\Theta$ imply
$\sup_t\|\hat\theta_t\|<\infty$ almost surely.
Thus, the comparison with the true parameter is preserved, and the
stability and tracking conclusions follow by localizing the proof
to deterministic parameter balls. The rate conclusions also carry
over when the corresponding parameter Lipschitz condition holds
uniformly on each such ball. Moreover, without constraints on the estimates or another mechanism
governing their selection, even exact global least-squares estimation
can lead to closed-loop instability. This differs from the stability results for the classical linear self-tuning
regulators analyzed in~\cite{guo1991,guo1995}. These linear results apply to
the specified estimators and feedback laws and do not cover every possible
selection from an arbitrary unregularized least-squares solution set. For a nonlinear example, let
\[
 b(z)=(1-4|z|)_+,\qquad
 f_\theta(y,u)=u-2y\,b(y-\theta),\qquad \theta\in\mathbb R,
\]
and take $\theta^*=0$, $y_0=1$,
$w_t\equiv0$, and $r_t\equiv0$. The model-difference inequality and
linear growth bound on the inverse map hold uniformly for $\theta\in\R$, and the true plant
satisfies the minimum-phase bound~\eqref{eq:mp}. Thus the assumptions hold on every
compact parameter set containing $\theta^*=0$; the unconstrained parameter domain
$\R$ itself is not compact.
Choose $\hat\theta_t=y_t$ and apply the
 input $u_t=2y_t$.
Then $y_t=2^t$. For every $s<t$,
$b(y_s-\hat\theta_t)=0$, so
\(
 Q_t(\hat\theta_t)=0
 =\min_{\theta\in\mathbb R}Q_t(\theta).
\)
Hence every estimate is an exact global least-squares solution, yet
$y_t=2^t\to\infty$.
\end{remark}
\section{Main Results}\label{sec:results}
\subsection{Lemmas on one-step prediction errors}\label{sec:prediction}
For~\eqref{eq:plant}, take $F_t(\theta)=f(\theta,Y_t,U_t)$ and $\zeta_t=1+\|Y_t\|+\|U_t\|$.
The prediction bounds are first established for a general predictable model family.
Assumption~\ref{ass:reach} is not used in this subsection.

Consider
\begin{equation}\label{eq:sequential-model}
 y_{t+1}=F_t(\vartheta^*)+w_{t+1},\qquad t\ge0,
\end{equation}
where $\vartheta^*$ belongs to a known nonempty compact set
$\Theta\subset\R^d$, $d\ge1$, of diameter at most $R>0$.
For each fixed parameter, the
one-step predictor uses only information available before
$y_{t+1}$ is observed. Suppose that, for a finite predictable $\zeta_t\ge1$,
\begin{equation}\label{eq:prediction-modulus}
 |F_t(\vartheta)-F_t(\vartheta')|
 \le \zeta_t\omega(\|\vartheta-\vartheta'\|),
 \qquad \vartheta,\vartheta'\in\Theta,
\end{equation}
where $\omega$ is finite, nondecreasing, $\omega(0)=0$, and
$\omega(h)\to0$ as $h\to0^+$.

At time $t$, form the cumulative least-squares criterion and select a finite
$\F_t$-measurable approximate minimizer:
\begin{align}
 Q_t(\vartheta)&=\sum_{s<t}[y_{s+1}-F_s(\vartheta)]^2,
 \label{eq:prediction-Q}\\
 Q_t(\hat\vartheta_t)&\le\min_{\vartheta\in\Theta}Q_t(\vartheta)
                         +e_t,\qquad \hat\vartheta_t\in\Theta.
 \label{eq:prediction-algorithm}
\end{align}
The tolerance $e_t\ge0$ is finite and predictable. The observed regressors
are held fixed when evaluating different parameter values. The analysis below uses only
the weaker comparison with the true parameter
\begin{equation}\label{eq:prediction-comparator}
 Q_t(\hat\vartheta_t)\le Q_t(\vartheta^*)+e_t=W_t+e_t.
\end{equation}

Define the prediction-error quantities
\begin{align}
 D_t(\vartheta)&=F_t(\vartheta)-F_t(\vartheta^*),\qquad
 \Delta_t=D_t(\hat\vartheta_t),\qquad
 \Reg_T=\sum_{t<T}\Delta_t^2,\label{eq:prediction-errors}\\
 S_t(\vartheta)&=\sum_{s<t}D_s(\vartheta)^2,\qquad
 M_t(\vartheta)=\sum_{s<t}D_s(\vartheta)w_{s+1},
 \label{eq:history-sums}\\
 B_T&=1+\max_{t<T}S_t(\hat\vartheta_t),\qquad
 M_T^*=\max_{t\le T}\sup_{\vartheta\in\Theta}|M_t(\vartheta)|,
 \label{eq:history}\\
 \mathcal E_T&=e+\sum_{t<T}\zeta_t^2,\qquad
 \mathcal P_T=\max_{t<T}\zeta_t^2,\qquad
 \bar e_T=\max_{t<T}e_t.\label{eq:prediction-energy}
\end{align}

\begin{lemma}[One-step prediction under a general modulus]
\label{lem:prediction}
Under~\eqref{eq:sequential-model}--\eqref{eq:prediction-modulus} and
Assumption~\ref{ass:noise}, every causal estimate sequence satisfying
\eqref{eq:prediction-comparator} obeys, almost surely,
\begin{equation}\label{eq:prediction-history-general}
 M_T^*=o(\mathcal E_T),\qquad
 B_T\le1+\bar e_T+2M_T^*,
\end{equation}
Moreover, there is a finite constant $C$, independent of $T\ge1$ and $0<h\le1$,
such that
\begin{equation}\label{eq:prediction-cover}
  {\quad
 \Reg_T\le C\{h^{-d}(B_T+\mathcal P_T)
                  +\omega(2Rh)^2\mathcal E_T\}.
 \quad}
\end{equation}
In particular, if $e_t=O(1)$ and $\zeta_t^2=O(1)$, then
\begin{equation}\label{eq:prediction-relative}
 B_T=o(T),\qquad \Reg_T=o(T),
 \quad\text{a.s.}
\end{equation}
\end{lemma}

\begin{lemma}[Prediction regret under parameter Lipschitz continuity]
\label{lem:prediction-rates}
In Lemma~\ref{lem:prediction}, suppose $\omega(h)=L_fh$.
Every causal estimate sequence satisfying~\eqref{eq:prediction-comparator}
obeys the following almost-sure bounds.

Under Assumption~\ref{ass:noise}, if
$\beta\ge (d+2)^2/[2(d+1)]$, then
\begin{equation}\label{eq:prediction-envelope-high}
 \Reg_T=O\!\left(
 \mathcal E_T^{d/(d+2)}
 [1+\bar e_T+\mathcal P_T+
       \sqrt{\mathcal P_T}\log\mathcal E_T]^{2/(d+2)}\right).
\end{equation}
If $2<\beta<(d+2)^2/[2(d+1)]$, put
$\tau=\beta(d^2+2d+2)/[2(d+2)]$.
For every fixed positive, eventually nondecreasing slowly varying
$\ell(\cdot)$ satisfying
\begin{equation}\label{eq:bounded-prediction-integral}
 \int_{e^e}^{\infty}\frac{dt}{t\ell(t)^\tau}<\infty,
\end{equation}
\begin{equation}\label{eq:prediction-envelope-low}
 \begin{aligned}
 \Reg_T={}&O\!\left(
 \mathcal E_T^{d/(d+2)}
       [1+\bar e_T+\mathcal P_T]^{2/(d+2)}\right)\\
 &+o\!\left(
 \mathcal E_T^{1-(\beta-1)/\tau}
 [1+\bar e_T+\mathcal P_T]^{(\beta-2)/(2\tau)}
 [\log(e+\bar e_T+\mathcal P_T)]^{2/\tau}
 \ell(\mathcal E_T)\right).
 \end{aligned}
\end{equation}

Moreover, for every $\beta>2$, the prediction regret also satisfies the following bound
in terms of $\mathcal E_{\beta,T}\triangleq e+\sum_{t<T}\zeta_t^\beta$:
for each fixed $\epsilon>0$,
\begin{equation}\label{eq:prediction-moment-final}
 \begin{aligned}
 \Reg_T=O\!\Bigg(\mathcal E_T^\epsilon\Bigg\{
 &\mathcal E_T^{d/(d+2)}
 \left[1+\bar e_T+
  \left(\frac{\mathcal E_{\beta,T}}{\mathcal E_T}\right)^{2/(\beta-2)}
 \right]^{2/(d+2)}\\
 &+\mathcal E_T^{\frac{d(\beta+2d-2)}{(d+2)\beta+2d(d+1)}}
 \mathcal E_{\beta,T}^{\frac{4d}{\beta[(d+2)\beta+2d(d+1)]}}
 +\mathcal E_{\beta,T}^{2/\beta}\Bigg\}\Bigg).
 \end{aligned}
\end{equation}
\end{lemma}

\begin{corollary}[Bounded prediction envelopes]
\label{cor:bounded-prediction}
Under the conditions of Lemma~\ref{lem:prediction-rates}, suppose
$\zeta_t=O(1)$ and $e_t=O(1)$ almost surely. Put
\[
 \beta_c(d)=\frac{(d+2)^2}{2(d+1)}.
\]
If $\beta\ge\beta_c(d)$, then
\begin{equation}\label{eq:bounded-prediction-high}
 \Reg_T=O\!\left(T^{d/(d+2)}(\log T)^{2/(d+2)}\right)
 \quad\text{a.s.}
\end{equation}
If $2<\beta<\beta_c(d)$, put $\kappa=(\beta-1)/\tau$,
with $\tau$ as in Lemma~\ref{lem:prediction-rates}.
For every $\ell$ specified in Lemma~\ref{lem:prediction-rates},
one has
\begin{equation}\label{eq:bounded-prediction-heavy}
 \Reg_T=o\!\left(T^{1-\kappa}\ell(T)\right)\quad\text{a.s.}
\end{equation}
In particular, one may take
$\ell(T)=(\log T)^{1/\tau}
(\log\log T)^{(1+\epsilon)/\tau}$ for any fixed $\epsilon>0$.
\end{corollary}

The integrability condition in~\eqref{eq:bounded-prediction-integral}
ensures the summability needed to derive almost-sure convergence rates
from finite-horizon probability bounds, similarly to the use of
summability functions in the weighted-LS analysis of~\cite{guo1996}.

\subsection{Closed-loop stability and asymptotically optimal tracking}\label{sec:closed-loop}

\subsubsection{Stability and optimal tracking}

Under Assumption~\ref{ass:noise}, the time-average squared tracking error satisfies
\[
 \liminf_{T\to\infty} \mathcal J_T\ge\sigma^2,
 \quad\text{a.s.}
\]
The following result establishes that asymptotically optimal tracking, i.e., $\mathcal{J}_T\to\sigma^2$, is achievable. 
\begin{theorem}[Sublinear closed-loop regret]
\label{thm:main}
Under Assumptions~\ref{ass:family}--\ref{ass:noise}, for every fixed
finite optimization tolerance $\epsopt\ge0$, the closed-loop system under the adaptive regulator
\eqref{eq:algorithm}--\eqref{eq:algorithmcontrol} satisfies
\begin{equation}\label{eq:regsublinear}
 \Reg_T=o(T)\qquad\text{a.s.}
\end{equation}
\end{theorem}

\begin{remark}\label{rem:regret-consequences}
As discussed in~\cite{guo1995}, $\Reg_T=o(T)$ implies, almost surely,
\begin{align}
 &\limsup_{T\to\infty}\frac1T\sum_{t<T}(y_t^2+u_t^2)
 <\infty,\label{eq:stability}\\
 &\max_{0\le t\le T}y_t^2+\max_{0\le t<T}u_t^2=o(T),
 \label{eq:peak}\\
 &\mathcal J_T\longrightarrow\sigma^2.\label{eq:optimaltracking}
\end{align}
Thus Theorem~\ref{thm:main} yields closed-loop stability and
asymptotically optimal tracking. Theorem~\ref{thm:main} 
remains valid if the conditional $\beta$th-moment bound in
Assumption~\ref{ass:noise} is replaced by
\(
 W_T/T\to\sigma^2,\text{a.s.},
\)
while retaining $\Et w_{t+1}=0$ and
$\Et w_{t+1}^2=\sigma^2$. In particular, centered i.i.d. innovations
independent of the past require only a finite second moment.
\end{remark}

\subsubsection{Convergence rates under Lipschitz continuity}
\label{sec:lipschitz-rates}

For $\beta>2$, define
\begin{equation}\label{eq:rate-kappa}
 \begin{aligned}
 \kappa_g&=\frac2{d+2},\qquad \kappa_p=1-\frac2\beta,\\
 \kappa_e&=\frac{2[\beta^2+2d(\beta-1)]}
 {\beta[(d+2)\beta+2d(d+1)]},\qquad
 \tau_e=\frac{\beta[(d+2)\beta+2d(d+1)]}{4(\beta+d)},\\
 \kappa_R&=\min\{\kappa_g,\kappa_p,\kappa_e\},\qquad
 \kappa_J=\min\{1/2,\kappa_R\}.
 \end{aligned}
\end{equation}
For each fixed $\epsilon>0$, put
\begin{equation}\label{eq:rate-log-budget}
 \begin{aligned}
 \Gamma_\epsilon(T)={}&
 \left(\frac{\log(e+T)}T\right)^{\kappa_g}
 +T^{-\kappa_p}(\log(e+T))^{2/\beta}
                        (\log\log(e^e+T))^{2(1+\epsilon)/\beta}\\
 &+T^{-\kappa_e}(\log(e+T))^{1/\tau_e}
                        (\log\log(e^e+T))^{(1+\epsilon)/\tau_e}.
 \end{aligned}
\end{equation}

\begin{theorem}[Convergence rates under parameter Lipschitz continuity]
\label{thm:rates}
Suppose Assumptions~\ref{ass:family}--\ref{ass:noise} hold, with
$\omega(h)=L_fh$ for a finite deterministic constant $L_f>0$, that is,
\begin{equation}\label{eq:paramlip}
 |f(\theta,Y,U)-f(\vartheta,Y,U)|
 \le L_f(1+\|Y\|+\|U\|)\|\theta-\vartheta\|.
\end{equation}
For every fixed finite optimization tolerance $\epsopt\ge0$, the closed-loop system under the adaptive regulator
\eqref{eq:algorithm}--\eqref{eq:algorithmcontrol} satisfies, for any $\epsilon>0$, 
\begin{equation}\label{eq:rate-moment}
 \begin{aligned}
 \frac{\Reg_T}{T}&=O(\Gamma_\epsilon(T)),\qquad
 |\mathcal J_T-\sigma^2|
 =O\!\left(\Gamma_\epsilon(T)
       +\sqrt{\frac{\log\log T}T}\right)
 \qquad\text{a.s.} 
 \end{aligned}
\end{equation}
\end{theorem}

\begin{remark}\label{rem:refinedrates}
If $\beta\geq6(d=1),\beta>4(d=2)$ or $\beta\geq d+2(d\geq 3)$,
then the first term in~\eqref{eq:rate-log-budget} determines the convergence rate, so
\[
 \Gamma_\epsilon(T)
 =O\!\left[\left(\frac{\log T}{T}\right)^{2/(d+2)}\right].
\]
For general $\beta>2$, the prediction and tracking rates are respectively
$O(T^{-\kappa_R+\epsilon})$ and $O(T^{-\kappa_J+\epsilon})$. More generally, the logarithmic factors multiplying $T^{-\kappa_p}$
and $T^{-\kappa_e}$ in~\eqref{eq:rate-log-budget} may be replaced,
respectively, by any fixed positive, eventually nondecreasing slowly
varying functions $\ell_p$ and $\ell_e$ satisfying
\begin{equation}\label{eq:rate-summability}
 \int_{e^e}^{\infty}\frac{dt}{t\ell_p(t)^{\beta/2}}<\infty,
 \qquad
 \int_{e^e}^{\infty}\frac{dt}{t\ell_e(t)^{\tau_e}}<\infty.
\end{equation}
Theorem~\ref{thm:rates} remains valid with these modified rate expressions; its constants
may also depend on the chosen functions.

\end{remark}
\begin{proposition}[Sharpness of the closed-loop regret rates]
\label{prop:closed-loop-sharpness}
For every $d\ge1$, each assertion below holds for a closed-loop system
satisfying the assumptions of Theorem~\ref{thm:rates}, with estimates selected as exact minimizers of the full cumulative
LS criterion. Each model is fixed independently of the time horizon and has
$p=1$, $q\le1$, a unique inverse map, conditional noise variance one, and the true plant
$y_{t+1}=u_t+\sin y_t+w_{t+1}$.
For the finite-moment statements, let $\ell$ be positive, eventually
nondecreasing and slowly varying, with
\begin{equation}\label{eq:closed-sharp-divergence}
 \int_{e^e}^{\infty}\frac{dt}{t\ell(t)^\tau}=\infty.
\end{equation}
\begin{enumerate}[label=\textup{(\roman*)}]
\item \emph{Noise peaks.} For every $\beta>2$ and every such $\ell$ with
$\tau=\beta/2$, the single-ReLU family
\[
 f(\theta,y,u)=u+\sin y+(\theta_1y-1)_+,\qquad
 \Theta=\{0\}\cup\{4^{-k}e_1:k\ge1\},
\]
with zero reference and a suitable noise sequence satisfies
\begin{equation}\label{eq:closed-sharp-heavy}
 \limsup_{T\to\infty}
 \frac{\Reg_T}{T^{1-\kappa_p}\ell(T)}=\infty
 \qquad\text{a.s.,}
\end{equation}
where $e_1$ is the first coordinate vector, and the estimate is selected from the NLS minimizers by
choosing the largest first coordinate.
\item \emph{Gaussian noise.} A closed-loop system with independent standard
Gaussian noise satisfies
\begin{equation}\label{eq:closed-sharp-gaussian}
 0<\limsup_{T\to\infty}
 \frac{\Reg_T}{T^{1-\kappa_g}(\log T)^{\kappa_g}}<\infty
 \qquad\text{a.s.}
\end{equation}
\item \emph{Intermediate finite moments.} For every $\beta>2$ with
$\kappa_e<\min\{\kappa_g,\kappa_p\}$ and every $\ell$ satisfying
\eqref{eq:closed-sharp-divergence} with $\tau=\tau_e$, a closed-loop system
satisfies~\eqref{eq:closed-sharp-heavy} with $\kappa_p$ replaced by $\kappa_e$.
\end{enumerate}
Therefore, the exponent $\kappa_R$, the Gaussian logarithmic factor, and the
finite-moment integrability boundaries in~\eqref{eq:rate-summability}
cannot, in general, be improved in the respective regimes where they determine
the regret rate in Theorem~\ref{thm:rates} for the regulator considered. The proof is given in Appendix~\ref{app:sharpness}.
\end{proposition}

\subsubsection{Robustness to unmodeled dynamics}
\label{sec:unmodeled}
Consider the perturbed system
\begin{equation}\label{eq:robust-plant}
 y_{t+1}=f(\theta^*,Y_t,U_t)+\eta_t+w_{t+1},
 \qquad
 |\eta_t|\le\delta(1+\|Y_t\|+\|U_t\|),
\end{equation}
where $\eta_t$ is finite and $\F_t$-measurable, and
$\delta\ge0$ is deterministic. The unmodeled term may, in particular,
have the form $\eta_t=\eta(Y_t,U_t)$.
Retain the same LS criterion~\eqref{eq:Q}, the parameter
estimator~\eqref{eq:algorithm} with fixed tolerance, and the CE control law
\eqref{eq:algorithmcontrol}. The definitions of $\Delta_t$ and $\Reg_T$
continue to refer to the nominal parameter $\theta^*$, and the
closed-loop tracking error satisfies
\begin{equation}\label{eq:robust-closed-loop}
 y_{t+1}-r_{t+1}=\eta_t-\Delta_t+w_{t+1}.
\end{equation}

\begin{theorem}[Stability and convergence rates with unmodeled dynamics]
\label{thm:robust}
Suppose Assumptions~\ref{ass:family} and~\ref{ass:noise} hold,
with the parameter Lipschitz bound~\eqref{eq:paramlip}.
Impose Assumption~\ref{ass:reach} on the perturbed trajectory,
with the same map $\psi$ and structural constants; when $q\ge1$,
the implicit constant in the $O$-bound~\eqref{eq:mp} is deterministic
and independent of $\delta$.
Then, there are deterministic constants $\delta_0>0$ (see \eqref{eq:robust-threshold} in the proof) and $C<\infty$,
depending on the fixed model, inverse-map, minimum-phase, and noise bounds, together with the initial history, such that, for every
$0\le\delta\le\delta_0$, every fixed
finite optimization tolerance $\epsopt\ge0$, the perturbed system \eqref{eq:robust-plant} under the adaptive regulator
\eqref{eq:algorithm}--\eqref{eq:algorithmcontrol} satisfies
\begin{equation}\label{eq:robust-stability}
 \limsup_{T\to\infty}\frac1T\sum_{t<T}(y_t^2+u_t^2)<\infty
 \qquad\text{a.s.}
\end{equation}
Moreover, with $\Gamma_\epsilon$ defined in~\eqref{eq:rate-log-budget},
for every fixed $\epsilon>0$,
\begin{equation}\label{eq:robust-moment-rates}
 \begin{aligned}
  \frac{\Reg_T}{T}
  &\le O(\Gamma_\epsilon(T))+C\delta^{4/(d+2)},
  \\|\mathcal J_T-\sigma^2|
  &\le O\!\left(\Gamma_\epsilon(T)
       +\sqrt{\frac{\log\log T}T}\right)
       +C\delta^{4/(d+2)}
       \qquad\text{a.s.}
 \end{aligned}
\end{equation}
where the constant $C$ in the robustness terms is independent of $T$ and
$\delta\in[0,\delta_0]$. The simplification under higher noise moments and the admissible slowly varying factors
in Remark \ref{rem:refinedrates} following Theorem \ref{thm:rates} apply here as well.
\end{theorem}

\begin{remark}[Adaptive control with neural network models]
\label{rem:nn-unmodeled}
A fixed finite-depth feedforward network $g_\vartheta$ with weights
and biases in a compact set and fixed globally Lipschitz activation
functions satisfies the following bounds, as shown by induction over the layers:
\[
 |g_\vartheta(Y)|\le C_g(1+\|Y\|),\qquad
 |g_\vartheta(Y)-g_{\vartheta'}(Y)|
 \le L_g(1+\|Y\|)\|\vartheta-\vartheta'\|.
\]
Such activations include ReLU $\max\{x,0\}$, leaky ReLU
$\max\{x,\alpha x\}$ with fixed $0<\alpha<1$, $\tanh x$,
the logistic sigmoid $(1+e^x)^{-1}$, and softplus
$\log(1+e^x)$. 
For a general nonlinear plant, the discrepancy between the true dynamics and their neural network approximation can be included in $\eta_t$ as an unmodeled term. The network architecture introduces a trade-off between approximation accuracy and closed-loop robustness: increasing the
depth or width of the network may reduce the approximation error
between the true dynamics and the nominal model, but it can also
increase the Lipschitz constants and the associated constants in the
model-inversion and stability conditions, thereby reducing the admissible
robustness threshold $\delta_0$ in \eqref{eq:robust-threshold}. Quantifying the dependence of this robustness threshold on network
architecture and approximation accuracy requires a more detailed
analysis of neural network approximation and closed-loop stability
and remains a subject for future research.
\end{remark}

\section{Proofs}\label{sec:proofs}
We first state two martingale estimates used in the closed-loop analysis. Their proofs are given in Appendix~\ref{app:offset-proof}.
\begin{lemma}\label{lem:proof-noise}
Under Assumption~\ref{ass:noise},
\begin{equation}\label{eq:noiselln}
 W_T/T\to\sigma^2,\qquad \max_{1\le s\le T}w_s^2=o(T)
 \quad\text{a.s.}
\end{equation}
For any predictable sequence $a_t$ with finite values, write
$A_T=\sum_{t<T}a_t^2$ and $N_T=\sum_{t<T}a_tw_{t+1}$. Then
\begin{equation}\label{eq:normalized-cross}
 N_T=o(A_T)\ \text{on }\{A_T\to\infty\},\qquad
 |N_T|=O(1+A_T)\quad\text{a.s.}
\end{equation}
Moreover, $N_T$ converges on $\{A_\infty<\infty\}$, and, for an
almost surely finite $K(\omega)$ independent of $T$,
\begin{equation}\label{eq:noise-tracking-transfer}
 \left|\frac1T\sum_{t<T}(a_t+w_{t+1})^2-\sigma^2\right|
 \le\frac{3A_T+K(\omega)}T+\frac{|W_T-T\sigma^2|}T.
\end{equation}
In addition, for any positive, eventually nondecreasing slowly varying $\ell_p$
satisfying the first condition in~\eqref{eq:rate-summability},
\begin{equation}\label{eq:noise-rates}
 |W_T-T\sigma^2|=
 \begin{cases}
 o(T^{2/\beta}\ell_p(T)),&2<\beta\le4,\\
 O(\sqrt{T\log\log(e^e+T)}),&\beta>4,
 \end{cases}\quad\text{a.s.}
\end{equation}
\end{lemma}

\begin{lemma}[Global and local martingale-offset bounds]\label{lem:offset}
Suppose Assumption~\ref{ass:noise} holds and the predictable model family
satisfies~\eqref{eq:prediction-modulus} with $\omega(h)=L_fh$.
Let the historical sums be defined by~\eqref{eq:history-sums}.

\textup{(i)} Put
\begin{equation}\label{eq:prediction-budgets}
 \Phi(x)=x^{2/\beta}(\log x)^3,\qquad x\ge e.
\end{equation}
Then
\begin{equation}\label{eq:prediction-offset}
 \sup_{\vartheta\in\Theta}
 \{2M_t(\vartheta)-\frac34S_t(\vartheta)\}
 =O(\Phi(\mathcal E_t))\quad\text{a.s.}
\end{equation}
If~\eqref{eq:prediction-comparator} holds, then
\begin{equation}\label{eq:prediction-history-coarse}
 B_T+\max_{1\le s\le T}w_s^2
 =O(1+\bar e_T+\Phi(\mathcal E_T))\quad\text{a.s.}
\end{equation}

\textup{(ii)} For a deterministic integer $N\ge3$, localize the
conditional $\beta$th moment to a deterministic bound and stop all
sums before $\mathcal E_{t+1}>N$ or $\zeta_t^2>P$, where $P\ge1$
is deterministic. For a deterministic nonempty compact
$\mathcal C\subset\Theta$ of diameter at most $a$, set
\begin{equation}\label{eq:regional-offset-definition}
 U_{\mathcal C}=\max_{t\le N}\sup_{\vartheta\in\mathcal C}
       (2M_t(\vartheta)-S_t(\vartheta)/2)_+,
 \qquad \mathcal E_{m,N}=e+\sum_{t<N}\zeta_t^m.
\end{equation}
All sums are formed with the stopped coefficients, and the envelope is set to zero after
stopping. Fix $\eta>0$, $A_0>0$, and $K<\infty$.
For $N^{-K}\le a\le\max\{1,R\}$, $N^\eta\le H\le N^K$,
and $2\le m\le b\le\beta$ with $b>2$,
\begin{equation}\label{eq:regional-polynomial}
 \E(U_{\mathcal C}-c_0H)_+\le C\left\{
 H^{1-b/2}+a^{md/(m+d)}
 \E[\mathcal E_{m,N}^{d/(m+d)}]
 H^{1-b/2-md/[2(m+d)]}\right\}+N^{-A_0}.
\end{equation}
For $2<b<d+2$, one also has
\begin{equation}\label{eq:regional-peak}
 \E(U_{\mathcal C}-c_0H)_+\le C\left\{
 H^{1-b/2}+P^{(b-2)/2}a^{bd/(d+2)}
 N^{1-b/(d+2)}H^{1-b(d+1)/(d+2)}\right\}+N^{-A_0}.
\end{equation}
The constants are independent of $N,P,a,H$ and $\mathcal C$; the constant $C$ may depend on the fixed exponents and the localized moment bound.
The same estimates hold after any further predictable stopping. 
\end{lemma}

\subsection{Proof of Lemma~\ref{lem:prediction}}
\label{sec:prediction-proofs}
\begin{proof}
\textit{Step 1: historical least-squares comparison.}
Choose $1/\beta<r<1/2$. Conditional-moment localization,
Markov's inequality, and the Borel--Cantelli lemma give
$w_t=o((t+2)^r)$ almost surely. On any fixed finite deterministic
net $\mathcal G\subset\Theta$, apply
\cite[Lemma~3.2]{huangguo1990doublearray} to the predictable
coefficients $D_s(\vartheta)$, $\vartheta\in\mathcal G$, with
$\varphi(t)=(t+2)^r$. Since
$S_T(\vartheta)\le\omega(R)^2\mathcal E_T$, it gives
\[
 \max_{\vartheta\in\mathcal G}\max_{t\le T}|M_t(\vartheta)|
 =O(\sqrt{\mathcal E_T}\log\mathcal E_T)
  +o(\sqrt{\mathcal E_T}\,T^r\log\log T)
 =o(\mathcal E_T)
 \quad\text{a.s.},
\]
where the last equality follows from $\mathcal E_T\ge T$ and $r<1/2$.
Work on a common event of probability one for the fixed finite nets of
radii $R2^{-k}$, $k\ge1$.
For a nearest net point $\vartheta'$,
\[
 \max_{t\le T}|M_t(\vartheta)-M_t(\vartheta')|
 \le\omega(R2^{-k})\sqrt{\mathcal E_TW_T}.
\]
Since $W_T=O(T)$ and $\mathcal E_T\ge T$, division by $\mathcal E_T$
and passage first to $T\to\infty$, then to $k\to\infty$, give
$M_T^*=o(\mathcal E_T)$. The exact identity
\begin{equation}\label{eq:lossidentity}
 Q_t(\vartheta)=W_t+S_t(\vartheta)-2M_t(\vartheta)
\end{equation}
and the least-squares comparison yield
$S_t(\hat\vartheta_t)\le e_t+2M_t(\hat\vartheta_t)$, proving
the bound on $B_T$.

\textit{Step 2: bounds on the cumulative one-step prediction errors.}
Partition $\Theta$ into cells of an $Rh$-net with at most
$3^dh^{-d}$ centers. In each occupied cell let $\tau_j<T$ be
the last time an estimate in that cell is selected. At any earlier visit $s$,
\[
 \Delta_s^2\le2D_s(\hat\vartheta_{\tau_j})^2
                  +2\zeta_s^2\omega(2Rh)^2.
\]
The first terms sum to at most
$2S_{\tau_j}(\hat\vartheta_{\tau_j})$ in that cell. The last visit,
which is not included in this historical error sum, contributes at most
$\omega(R)^2\mathcal P_T$. Summing over cells proves
\eqref{eq:prediction-cover}.

If $e_t=O(1)$ and $\zeta_t^2=O(1)$, then
$\mathcal E_T\asymp T$ and $B_T+\mathcal P_T=o(\mathcal E_T)$.
Divide \eqref{eq:prediction-cover} by $\mathcal E_T$, let
$T\to\infty$ at fixed $h$, and then let $h\to0$. This proves~\eqref{eq:prediction-relative}.
\end{proof}

\subsection{Proof of Lemma~\ref{lem:prediction-rates}}\label{sec:prediction-rates-proof}
\begin{proof}
To refine the bound in Lemma~\ref{lem:prediction}, we estimate the sum of
the historical prediction errors over individual parameter regions,
in place of using the global bound $B_T$ for every region.

Localize the conditional $\beta$th moment, and stop before
$\mathcal E_{t+1}>N$, $\zeta_t^2>P$, or $e_t>e_0$, for deterministic
$N\ge e$, $P\ge1$, and $e_0\ge0$. All subsequent summands are set to zero;
the estimates selected before stopping are computed from their complete observation histories. Write
$\mathcal E_{m,N}=e+\sum_{t<N}\zeta_t^m$, and localize it on dyadic
budgets up to $e+N^{m/2}$ when needed.

\medskip\noindent\textit{Step 1: bound the contribution of each parameter region.}
For each region, use $U_\mathcal C$ from~\eqref{eq:regional-offset-definition}.
The least-squares comparison gives $S_t(\hat\vartheta_t)\le2U_\mathcal C+2e_0$ whenever
$\hat\vartheta_t\in \mathcal C$. Applying the last-visit argument used to prove
\eqref{eq:prediction-cover} in Lemma~\ref{lem:prediction}, with a cover of resolution $h$, gives
\[
 \Reg_N\le C\left[Nh^2+h^{-d}(e_0+P)+\sum_\mathcal C U_\mathcal C\right].
\]
For a more precise bound, group the errors according to
$a<|\Delta_t|/\zeta_t\le2a$, at dyadic scales $h\le a\le L_fR$,
and use a sufficiently fine cover of diameter proportional to $a$
for each group. Errors $|\Delta_t|$ below $h\zeta_t$ contribute at most $Nh^2$, while the contribution of earlier visits is bounded by $C(U_\mathcal C+e_0)$ per cell, by the definition of $U_\mathcal{ C}$. Moreover, the last visits are
distinct within each group. Hence, by $|\Delta_t|^2\leq 4a^2\zeta_t^2$, for $m\ge2$, their total contribution is bounded by \begin{equation}\label{eq:unified-last-visits}
 J_N\triangleq C\sum_{a\in\mathcal A}v_a
 \min\{a^{-d}P,N,
       \mathcal E_{m,N}^{2/m}a^{-d(1-2/m)}\},
\end{equation}
where $\mathcal A$ is the set of scales, $v_a=a^2$, and the term involving $m$ follows from H\"older's inequality.
For a single cover take $\mathcal A=\{h\}$ and $v_h=1$.

Decomposing each $U_\mathcal C$ at the threshold $c_0H_a$, with $H_a\ge1$, and applying
Markov's inequality only to the sum of excesses yields
\begin{equation}\label{eq:unified-prediction-tail}
 \begin{aligned}
 &\PP\!\left\{\Reg_N>C\left[
 Nh^2+e_0h^{-d}+J_N+\sum_{a\in\mathcal A}a^{-d}H_a+x\right]\right\}\\
 &\hspace{8mm}\le\frac Cx\sum_{a\in\mathcal A}a^{-d}
    \sup_{\operatorname{diam}(\mathcal C)\le a}\E(U_\mathcal C-c_0H_a)_+.
 \end{aligned}
\end{equation}
Lemma~\ref{lem:offset} provides the regional excess bounds
\eqref{eq:regional-polynomial}--\eqref{eq:regional-peak}.

On the full parameter set, \eqref{eq:regional-polynomial} with
$b=\beta$ and
$H=\mathcal E_T^{\epsilon'}
\mathcal E_{m,T}^{2d/[\beta(m+d)+md]}$ also gives, by Markov's
inequality and dyadic summation,
\begin{equation}\label{eq:prediction-history-rate}
 B_T=O\!\left(1+\bar e_T+\mathcal E_T^\epsilon
 \mathcal E_{m,T}^{\frac{2d}{\beta(m+d)+md}}\right)
 \quad\text{a.s.},\qquad 2\le m\le\beta.
\end{equation}

\medskip\noindent\textit{Step 2: select the covering resolutions and error thresholds.}
For~\eqref{eq:prediction-envelope-high}, take
$b=(d+2)^2/[2(d+1)]$, localize $1+\bar e+P$ at $A\le N$, and use
one cover with
\[
 H=C_0(A+\sqrt P\log N),\qquad h=(H/N)^{1/(d+2)}.
\]
Equation~\eqref{eq:unified-peak-budget} gives
$\sum q_t(Cs)^b\le CP^{(b-2)/2}s$ for $s\ge H$.
Apply~\eqref{eq:offset-truncation} with $B=K\sqrt P$.
The discarded absolute products have total conditional mean
at most $CK^{1-b}P^{-1/2}s$, which is chosen smaller than $s/100$;
their centered sum has variance budget $CK^{2-b}s$ and
$b$th-moment budget $CP^{(b-2)/2}s$.
The probability of an excursion of the centered sum is therefore bounded by
$C[s^{-b/2}+P^{(b-2)/2}s^{1-b}]$.
The exponential probability bound remains summable after taking the union over the net because
$H\ge C_0\sqrt P\log N$. Integrating and using
\eqref{eq:unified-prediction-tail} bounds the probability of
$\Reg_N>CN^{d/(d+2)}H^{2/(d+2)}$ by
$C[H^{-b/2}+P^{(b-2)/2}H^{1-b}]+N^{-A_0}$.
At $N=2^k$, its sum over dyadic $P\le A\le N$ is at most
\(
 C[1+\log(k+1)](k^{-b/2}+k^{1-b})+Ck^2 2^{-kA_0},
\)
which is summable since $b>2$.

For~\eqref{eq:prediction-envelope-low}, put
$\kappa=(\beta-1)/\tau$, $\gamma=(\beta-2)/(2\tau)$, and choose
\[
 h=\max\{(A/N)^{1/(d+2)},(N^{-\kappa}A^\gamma z)^{1/2}\},
 \qquad H=Nh^{d+2},\qquad z\ge1.
\]
Here the threshold $H$ grows polynomially with the horizon. Equations~\eqref{eq:regional-peak}
and~\eqref{eq:unified-prediction-tail} give
\[
 \PP\{\Reg_N>C[N^{d/(d+2)}A^{2/(d+2)}
                  +N^{1-\kappa}A^\gamma z]\}
 \le Cz^{-\tau}+CN^{-c},
\]
since the residual term is at most
$CA^{(\beta-2)/2}N^{1-\beta}h^{-2\tau}$.
Set $z=\varepsilon[\log(e+A)]^{2/\tau}\ell(N)$.
Condition~\eqref{eq:bounded-prediction-integral} ensures that the sum over
dyadic $A,N$ is finite; rational rescaling in $\varepsilon>0$ proves
the little-$o$ assertion.

Finally, for~\eqref{eq:prediction-moment-final}, use the error groups
and~\eqref{eq:regional-polynomial} with $m=b=\beta$.
At a dyadic budget $\mathcal E_{\beta,N}\le V$, choose
$H_a=N^\eta(1+a^{\beta b_0}V^{b_0})$, where
$b_0=2d/[\beta(\beta+2d)]$ and $\eta>0$ is arbitrarily small.
The normalized excess bound obtained by division by $H_a$ decays polynomially; summing
\eqref{eq:unified-prediction-tail} gives
\[
 \Reg_T=O\!\left(\mathcal E_T^\epsilon
 \left\{\mathcal E_T h^2+(1+\bar e_T)h^{-d}
 +h^{\beta b_0-d}\mathcal E_{\beta,T}^{b_0}
 +\mathcal E_{\beta,T}^{2/\beta}
 [1+h^{2-d(1-2/\beta)}]\right\}\right).
\]
Logarithmic factors arising from summation over the scales are absorbed into $\mathcal E_T^\epsilon$.

Use
\(
 \mathcal E_{\beta,T}^{2/\beta}h^{2-d(1-2/\beta)}
 \le C\mathcal E_T h^2+
 C\left(\frac{\mathcal E_{\beta,T}}{\mathcal E_T}\right)^{2/(\beta-2)}h^{-d}
\)
and balance each decreasing power of $h$ against
$\mathcal E_T h^2$. This proves~\eqref{eq:prediction-moment-final}.
If a covering resolution exceeds one or $1+e_0+P>N$, the bound
$\Reg_T\le C\mathcal E_T$ suffices. The Borel--Cantelli lemma, countable
localization, and monotonicity extend the dyadic bounds to all $T$.
\end{proof}

\subsection{Proof of Corollary~\ref{cor:bounded-prediction}}
\begin{proof}
Here $\mathcal E_T\asymp T$ and $1+\bar e_T+\mathcal P_T=O(1)$.
Substitute these bounds in~\eqref{eq:prediction-envelope-high}
or~\eqref{eq:prediction-envelope-low}.
In the latter case $1-\kappa>d/(d+2)$, so the first term is
absorbed by the stated little-$o$ bound. Slow variation permits
replacing $\ell(\mathcal E_T)$ by $\ell(T)$.
\end{proof}

\subsection{Proof of Theorem~\ref{thm:main}}
\label{sec:main-proof}\label{sec:stability-proof}
\begin{proof}
\textit{Step 1: bound the regressors using an exponentially weighted signal energy.}
Choose $\lambda$ from the minimum-phase condition~\eqref{eq:mp}; when $q=0$,
any fixed $\lambda\in(0,1)$ suffices. Following the signal-energy argument in~\cite{guo1995}, define
\begin{equation}\label{eq:memory}
 V_t^2=1+\sum_{s=0}^t\lambda^{t-s}(y_s^2+w_s^2),\qquad
 H_T=\sum_{t<T}V_t^2,\qquad \rho=\sqrt\lambda.
\end{equation}
Finite memory and the minimum-phase bound, with the finite initial
segment absorbed into $C$, give $1+\|Y_t\|+\|U_t^-\|\le CV_t$.
The selected inverse map yields
\begin{equation}\label{eq:regbound}
 \zeta_t\triangleq1+\|Y_t\|+\|U_t\|\le c_VV_t,\qquad
 \mathcal E_T\le C(1+H_T).
\end{equation}

The closed-loop identity gives
\begin{align}
 V_{t+1}^2
 &\le\lambda V_t^2+3\Delta_t^2+C(1+w_{t+1}^2),
 \label{eq:Vsquare}\\
 V_{t+1}
 &\le\rho V_t+A|\Delta_t|+C(1+|w_{t+1}|),
 \label{eq:Vrec}\\
 H_T&\le C(T+\Reg_T+W_T).
 \label{eq:energy}
\end{align}
where the last inequality follows by summing the first.

\textit{Step 2: bound signal peaks in terms of historical prediction errors.}
Fix a sample path and a horizon $T$. Choose $\kappa>0$ with
$\rho+A\kappa\triangleq\rho_0<1$, and a finite parameter cover independent
of $T$ such that models in the same cell differ by at most
$\kappa V_t/2$, as ensured by Assumption~\ref{ass:family}. A time is exceptional if
$|\Delta_t|>\kappa V_t$ and $V_t>2\sqrt{B_T}/\kappa$.
There is at most one exceptional time in each cell: two such
times $i<j$ would give
\[
 |D_i(\hat\theta_j)|
 \ge|\Delta_i|-\kappa V_i/2>\sqrt{B_T},
\]
contradicting $S_j(\hat\theta_j)\le B_T$.

Above a sufficiently large multiple of
$\sqrt{B_T}+\max_{s\le T}|w_s|$ that exceeds $V_0$,
nonexceptional steps cannot increase $V_t$, by~\eqref{eq:Vrec}.
At the remaining steps, the signal energy increases by at most a fixed factor, since
$|\Delta_t|\le CV_t$. The fixed parameter cover bounds the number of exceptional
steps, giving
\begin{equation}\label{eq:detpeak}
 \max_{t\le T}V_t^2+\mathcal P_T
 =O\left(B_T+\max_{1\le s\le T}w_s^2\right).
\end{equation}

\textit{Step 3: establish the total signal-energy bound.}
Lemma~\ref{lem:prediction}, with $e_t=\epsopt$, and
\eqref{eq:noiselln} give $B_T+\max_{s\le T}w_s^2=o(H_T)$,
since $\mathcal E_T\le C(1+H_T)$ and $H_T\ge T$.
Thus $\mathcal P_T=o(H_T)$ by~\eqref{eq:detpeak}.
Divide~\eqref{eq:prediction-cover} by $H_T$, let $T\to\infty$
at fixed $h$, then let $h\to0$. This gives $\Reg_T=o(H_T)$,
and~\eqref{eq:energy} yields
\begin{equation}\label{eq:energy-closure}
 H_T\le O(T)+o(H_T),\qquad H_T=O(T),\qquad \Reg_T=o(T).
\end{equation}
This proves the theorem.
\end{proof}

\subsection{Proof of Theorem~\ref{thm:rates}}\label{sec:rates-proof}
\begin{proof}
The proof of Theorem~\ref{thm:main} gives $\mathcal E_T=O(T)$,
and Remark~\ref{rem:regret-consequences} gives the tracking error bound. We first refine the
weighted signal-energy estimate to bound higher moments of the regressor envelope,
and then apply the regional martingale-offset and prediction estimates in
Lemmas~\ref{lem:offset} and~\ref{lem:prediction-rates}.

\medskip\noindent\textit{Step 1: bound higher moments of the regressor envelope.}
The fixed-cover argument in Theorem~\ref{thm:main} gives
\begin{equation}\label{eq:signal-beta-energy}
 \mathcal E_{m,T}
 \le C\left(1+T+\sum_{s\le T}|w_s|^m+B_T^{m/2}\right),
 \qquad m\ge2.
\end{equation}
Indeed, stop before the historical prediction error of the selected estimate exceeds $D$.
Call a time $t$ exceptional when $|\Delta_t|>\kappa V_t$, with
$\rho+A\kappa=\rho_0<1$. In each fixed sufficiently fine parameter
cell, the occurrence of a subsequent exceptional visit bounds the sum of the squared errors at preceding exceptional
visits by $4D$. Mark the first exceptional visit that takes this sum
above $4D$; no later exceptional visit in the cell is possible.
Set $b_t=|\Delta_t|$ at unmarked exceptional times and zero otherwise.
Then $\sum b_t^2\le CD$. At the bounded number of marked times,
$|\Delta_t|\le CV_t$ allows only a fixed multiplicative increase in the signal bound.
Iteration of~\eqref{eq:Vrec} therefore yields the predictable bound
\begin{equation}\label{eq:predictable-memory-split}
 \begin{aligned}
 \zeta_t&\le C(Z_t+E_t),\qquad
 Z_t=1+\sum_{s=1}^t\rho_0^{t-s}|w_s|,\\
 E_t&=\sum_{s<t}\rho_0^{t-1-s}b_s,\qquad
 \sum_{t<T}E_t^m\le CD^{m/2}.
 \end{aligned}
\end{equation}
The stable-convolution inequality, with $D=B_T$ pathwise, proves
\eqref{eq:signal-beta-energy}.

Equation~\eqref{eq:prediction-history-rate}, with $m=2$, gives
$B_T=O(T^{2d/[\beta(d+2)+2d]+\eta})$ for every $\eta>0$.
Choose $2d/[\beta(d+2)+2d]<a_0<2/\beta$.
At a dyadic horizon $N$, stop before any of the predictable bounds on
energy, historical prediction error, regressor growth, or conditional moments
\[
 \mathcal E_{t+1}\le C_0N,\quad
 S_t(\hat\theta_t)\le KN^{a_0},\quad
 \zeta_t\le c_VV_t,\quad \Et|w_{t+1}|^\beta\le j
\]
fail. Countable localization over deterministic $C_0,K,c_V,j$
removes these stops eventually almost surely. Up to this stopping time,
\eqref{eq:predictable-memory-split} and $a_0\beta/2<1$ give
\begin{equation}\label{eq:stopped-envelope-moment}
 \E\mathcal E_{\beta,N}\le C(N+N^{a_0\beta/2})\le CN.
\end{equation}

\medskip\noindent\textit{Step 2: historical error thresholds with polynomial growth.}
Use the normalized-error groups in
\eqref{eq:unified-prediction-tail} with $H=Nh^{d+2}\ge N^\eta$.
Lemma~\ref{lem:offset}\textup{(ii)}, with $b=m=\beta$,
and~\eqref{eq:stopped-envelope-moment} give
\begin{equation}\label{eq:closed-regret-tail}
 \PP\{\Reg_N>CNh^2+CJ_h\}
 \le C\left[
 H^{-\beta/2}
 +h^{\frac{\beta d}{\beta+d}}N^{\frac d{\beta+d}}
 H^{-\frac\beta2-\frac{\beta d}{2(\beta+d)}}\right]+N^{-A_0},
\end{equation}
where $J_h$ is the contribution of the last visits in
\eqref{eq:unified-last-visits} and $A_0>0$ is arbitrary.
Both regional sums are dominated by their finest scale, since
$\beta d/(\beta+d)<d$. Distinct last visits, H\"older's inequality,
and~\eqref{eq:stopped-envelope-moment} also give
\begin{equation}\label{eq:closed-last-moment}
 \E J_h^{\beta/2}\le CN
 \begin{cases}
 1,&2-d(1-2/\beta)>0,\\
 [\log(e/h)]^{\beta/2},&2-d(1-2/\beta)=0,\\
 h^{\beta-d(\beta-2)/2},&2-d(1-2/\beta)<0.
 \end{cases}
\end{equation}
If $\kappa_e<\min\{\kappa_g,\kappa_p\}$, take
$h^2=N^{-\kappa_e}\ell_e(N)$: the second probability bound in
\eqref{eq:closed-regret-tail} is $C\ell_e(N)^{-\tau_e}$,
and the remaining probability bounds decay polynomially.
If $\kappa_p<\kappa_g$ and $\kappa_p\le\kappa_e$, take
$h^2=N^{-\kappa_p}\ell_p(N)$: here $2-d(1-2/\beta)>0$,
and~\eqref{eq:closed-last-moment} gives
$C\ell_p(N)^{-\beta/2}$. At equality $\kappa_p=\kappa_e$,
$\tau_e>\beta/2$ also makes the regional term summable.
Thus~\eqref{eq:rate-summability} proves the required bounds whenever
$\kappa_R<\kappa_g$.

\medskip\noindent\textit{Step 3: historical error thresholds with logarithmic growth.}
The remaining cases are $d\ge3$, $\beta\ge d+2$;
$d=2$, $\beta\ge4$; and $d=1$, $\beta\ge6$.
Put $h=(H/N)^{1/(d+2)}$. Since the local excess bounds in
Lemma~\ref{lem:offset} require $H\ge N^\eta$, we instead use
the truncation estimate~\eqref{eq:offset-truncation} directly
at logarithmic thresholds. The required moment bounds are verified
below using the closed-loop estimates.

First consider $\beta=d+2\ge4$. For $1\le\alpha<\beta$,
$\min(x^\beta,x^\alpha)$ has a convex equivalent. Applied to the
noise convolution in~\eqref{eq:predictable-memory-split}, Jensen's
inequality gives, for fixed $c>0$,
\[
 \sum_{t<N}\min\{(chZ_t)^\beta,(chZ_t)^\alpha\}
 \le C\left[N(ch)^\beta+
 \sum_{s\le N}\min\{(Cch|w_s|)^\beta,(Cch|w_s|)^\alpha\}\right].
\]
Since $Nh^\beta=H$, the conditional means, variances, and
$\beta/\alpha$th-moment sums on the right are $O(H)$.
For $\beta/\alpha\ge2$, the maximal moment inequality bounds
the probability of an excursion of order $H$ by
$C[H^{-\beta/(2\alpha)}+H^{1-\beta/\alpha}]$.
Take $\alpha=1,2,b$ with $2<b<\beta/2$ if $\beta>4$,
and $\alpha=1,2$ if $\beta=4$.
These probabilities are dyadically summable with $H=C\log N$
in the first case, and $H=C\ell_p(N)^2$ in the second.
The latter choice has $\sum_kH(2^k)^{-1}<\infty$ and
$H/\log N\to\infty$ by~\eqref{eq:rate-summability}.

Consequently, outside events with summable probabilities, for $r=1,2,b$
(with $b=2$ at $\beta=4$),
\[
 \sum_{t<N}\Et\!\left[(Ch\zeta_t|w_{t+1}|)^r
       \mathbf1_{\{Ch\zeta_t|w_{t+1}|>B_0\}}\right]
 \le C_jB_0^{r-\beta}H+o(H).
\]
The remainder in~\eqref{eq:predictable-memory-split} contributes
$O(h^\beta N^{a_0\beta/2})=o(H)$.
Choose $B_0$ large enough to make the discarded mean a small
multiple of $H$, and stop before these predictable budgets fail.
The moment term in~\eqref{eq:offset-truncation} is then
$C(H^{-b/2}+H^{1-b})$.
At scale $a=xh$, take $H_a=xH$ and $B_a=xB_0$:
the discarded product is $x$ times that at scale $h$, so a single
event controls all scales. The retained-product exponent is at least
$cH_a/(1+B_a)\ge c_jH$, which ensures summability after taking unions over the nets and energy intervals.
Equation~\eqref{eq:scalar-offset-tail} bounds the scalar-offset
contribution by $CH^{-\beta/2}$ in probability, and
\[
 \sum_{a\ge h}a^{-d}H_a\le Ch^{-d}H=CNh^2\qquad(d>1).
\]
This bounds the total contribution of all earlier visits. If $d\ge2$ and $\beta>d+2$,
the argument simplifies: use $m=d+2<\beta$,
localize $\mathcal E_{m,N}=O(N)$ from
\eqref{eq:predictable-memory-split} and the martingale strong law,
and take $H=C\log N$. The moment bounds for the discarded products
are then directly $O(H)$.

For $d=1$, $\beta\ge6$, use $\mathcal E_{4,N}=O(N)$,
$H=C\log N$, and $H_a=(a/h)^{3/4}H$,
$B_a=(a/h)^{3/4}B_0$. The discarded mean is absorbed by choosing
$B_0$ large, and~\eqref{eq:offset-truncation} with $b=4$ gives
\[
 C\left\{\frac{(B_a^{-2}a^4N)^2}{H_a^4}
              +\frac{a^4N}{H_a^4}\right\}
 =C(a^2H^{-2}+aH^{-3}).
\]
The sum of these bounds over the scales is summable along dyadic horizons, and
$\sum a^{-1}H_a\le Ch^{-1}H$; the same retained-product bound applies.

It remains to bound the contribution of the last visits. For $d>2$,
$J_h\le C\mathcal E_N^{1-2/d}\mathcal P_N^{2/d}$ and
$\mathcal P_N\le C(1+\max_{s\le N}w_s^2+KN^{a_0})$.
A union bound at
$\max w_s^2=N^{2/(d+2)}H^{d/(d+2)}$ gives
$\PP\{J_h>CNh^2\}\le CH^{-d/2}$.
For $d=2$, disjoint last-visit sets give
$J_h\le C\mathcal E_{4,N}^{1/2}$: this is $O(\sqrt N)$ if
$\beta>4$, while at $\beta=4$ the probability of exceeding
$\sqrt{NH}$ is $C/H$ by~\eqref{eq:stopped-envelope-moment}.
For $d=1$, $J_h\le C\mathcal P_N$, whose probability of exceeding
$CN^{1/3}H^{2/3}$ is at most $CH^{-2}$.
The remainder $KN^{a_0}$ is of lower order in each case because $a_0<2/\beta$.
All these probabilities are summable at the chosen $H$, so
\begin{equation}\label{eq:log-level-prediction}
 \Reg_N\le CN^{d/(d+2)}H^{2/(d+2)}
\end{equation}
eventually almost surely.

Steps~2--3 apply to every $\beta>2$. Countable localization and
monotonicity extend the dyadic bounds to all horizons, giving
\eqref{eq:rate-moment} with the stated slowly varying factors or those in
\eqref{eq:rate-summability}.
\end{proof}

\subsection{Proof of Theorem~\ref{thm:robust}}\label{sec:robust-proof}
\begin{proof}
We adapt the signal-peak and energy arguments used in Theorem~\ref{thm:main}.
Here $C$ denotes deterministic structural constants independent of
$\delta$, while $K(\omega)$ may absorb finite initial contributions
and the fixed optimization tolerance. Use
$F_t(\theta)=f(\theta,Y_t,U_t)$ and $\Phi$ from
\eqref{eq:prediction-budgets}. 

\medskip\noindent\textit{Step 1: establish the signal-energy bound in the presence of unmodeled dynamics.}
The least-squares residual identity becomes
\begin{equation}\label{eq:robust-loss}
 Q_t(\theta)-Q_t(\theta^*)
 =S_t(\theta)-2M_t(\theta)-2\sum_{s<t}\eta_sD_s(\theta).
\end{equation}
Lemma~\ref{lem:offset}\textup{(i)} gives~\eqref{eq:prediction-offset}
also on the perturbed trajectory, since this estimate uses only predictability and
the noise assumptions. Combining the least-squares comparison with
$2|\sum\eta_sD_s|\le S_t/8+8\delta^2\mathcal E_t$ gives
\begin{equation}\label{eq:robust-history}
 B_T\le C\delta^2\mathcal E_T+K(\omega)\Phi(\mathcal E_T).
\end{equation}
The assumptions give $\zeta_t\le c_VV_t$ eventually, with deterministic
$c_V$ independent of $\delta$, and
\[
 V_{t+1}\le(\sqrt\lambda+c_V\delta)V_t
                  +|\Delta_t|+C(1+|w_{t+1}|).
\]
For $\delta\le\min\{1,(1-\sqrt\lambda)/(2c_V)\}$ the contraction
factor is at most $(1+\sqrt\lambda)/2<1$.
Thus the signal-peak argument based on a fixed parameter cover in Theorem~\ref{thm:main},
with the initial segment absorbed into $K(\omega)$, gives
\begin{equation}\label{eq:robust-history-peak}
 B_T+\mathcal P_T+\max_{t\le T}V_t^2+\max_{s\le T}w_s^2
 \le C\delta^2\mathcal E_T+K(\omega)\Phi(\mathcal E_T).
\end{equation}
The pathwise comparison~\eqref{eq:prediction-cover} remains valid.
Optimizing the covering resolution, or using $\Reg_T\le(L_fR)^2\mathcal E_T$
if that resolution exceeds one, yields
\begin{equation}\label{eq:robust-prediction-budget}
 \Reg_T\le C_R\delta^{4/(d+2)}\mathcal E_T
 +K(\omega)\mathcal E_T^{d/(d+2)}\Phi(\mathcal E_T)^{2/(d+2)},
\end{equation}
with deterministic $C_R$. Summing the weighted signal-energy estimates and using
$W_T/T\to\sigma^2$ gives
\begin{equation}\label{eq:robust-energy}
 \mathcal E_T\le C_0T+C_1\Reg_T+C_2\delta^2\mathcal E_T+K(\omega)
\end{equation}
eventually almost surely, with deterministic $C_0,C_1,C_2>0$.
Choose
\begin{equation}\label{eq:robust-threshold}
 \delta_0=\min\left\{
 1,\frac{1-\sqrt\lambda}{2c_V},
 (8C_1C_R)^{-(d+2)/4},(8C_2)^{-1/2}\right\}.
\end{equation}
The terms involving $\delta$ are then bounded by one quarter
of the left-hand side of~\eqref{eq:robust-energy}. Since $\Phi(x)=o(x)$,
the remaining sublinear term is absorbed as well. Consequently
$\mathcal E_T\le CT$ eventually, with deterministic $C$ independent
of $\delta$, proving~\eqref{eq:robust-stability}.

\medskip\noindent\textit{Step 2: prediction and tracking bounds.}
Apply~\eqref{eq:noise-tracking-transfer} with
$a_t=\eta_t-\Delta_t$. Since
$\sum_{t<T}a_t^2\le2\Reg_T+2\delta^2\mathcal E_T$,
\begin{equation}\label{eq:robust-tracking-transfer}
 |\mathcal J_T-\sigma^2|
 \le C\frac{\Reg_T+\delta^2\mathcal E_T}{T}
       +\frac{K(\omega)}T+\frac{|W_T-T\sigma^2|}{T}.
\end{equation}
Equations~\eqref{eq:robust-prediction-budget} and
\eqref{eq:noiselln}, together with $\mathcal E_T\le CT$ and
$\delta^2\le\delta^{4/(d+2)}$, imply, for a deterministic
$C_*<\infty$ independent of $\delta$,
\begin{equation}\label{eq:robust-asymptotic-bounds}
 \limsup_{T\to\infty}
 \max\left\{\frac{\Reg_T}{T},|\mathcal J_T-\sigma^2|\right\}
 \le C_*\delta^{4/(d+2)}\quad\text{a.s.}
\end{equation}
For each fixed $\delta>0$, enlarging $C_*$ to $2C_*+1$ bounds
both quantities by $C\delta^{4/(d+2)}$ eventually, which implies
\eqref{eq:robust-moment-rates}. For $\delta=0$, those rates and
their slowly varying refinements follow directly from
Theorem~\ref{thm:rates}. This proves the stated bounds for every
fixed $\delta\in[0,\delta_0]$ with a common deterministic constant in the robustness bound.
\end{proof}

\section{Conclusion}\label{sec:conclusion}
This paper has studied a least-squares-based adaptive regulator for general nonlinear stochastic systems. Compactness of the parameter space and the prescribed continuity bound in the parameters allow the historical least-squares comparison to be used to bound the cumulative one-step prediction error along the closed-loop trajectory. Together with model inversion and the minimum-phase condition, these bounds establish time-average closed-loop stability, sublinear growth of squared signal peaks, and asymptotically optimal tracking. The results allow a fixed optimization tolerance and do not require convergence of the parameter estimates. For Lipschitz parameterizations, explicit prediction and tracking convergence rates are obtained under finite-moment noise. Closed-loop examples show that the derived regret convergence rates cannot, in general, be improved for the regulator considered. The robustness analysis further shows that small unmodeled dynamics preserve time-average stability and contribute a nonvanishing term to the tracking error bound.

Further work includes identifying classes of nonlinear systems for which the approximate global least-squares requirement can be satisfied with efficient online computation. Extensions to adaptive control with integrated offline and online learning, time-varying parameters, and constrained control inputs also merit investigation.

\section*{Acknowledgment}

The author would like to thank his supervisor, Prof. Lei Guo, for suggesting
this research problem and for his valuable guidance and discussions throughout the study.

\appendix
\section{Lower bounds on the closed-loop regret}
\label{app:sharpness}
\subsection{Proof of Proposition~\ref{prop:closed-loop-sharpness}}
\begin{proof}
Set $\theta^*=0$ and write the difference in full cumulative
least-squares losses as
\[
 q_t(\theta)=Q_t(\theta)-Q_t(0)
 =\sum_{s<t}\{D_s(\theta)^2-2D_s(\theta)w_{s+1}\}.
\]
\textit{(i) A single ReLU and rare noise peaks.}
Write $h_\theta(y)=(\theta_1y-1)_+$. Since
\[
 0\le h_\theta(y)\le |y|/4,\qquad
 |h_\theta(y)-h_\vartheta(y)|\le |y|\|\theta-\vartheta\|,
\]
this compact model family satisfies the required parameter Lipschitz bound
and admits the unique inverse with linear growth $r-\sin y-h_\theta(y)$.
The true plant satisfies $u_{t-1}^2\le3(y_t^2+w_t^2+1)$.
Take $y_0=0$ and a zero reference signal, and use the tie-breaking rule
specified in the proposition.

We first introduce two elementary constructions that will also be used
in part~\textup{(iii)}.
If~\eqref{eq:closed-sharp-divergence} holds, choose
$x_{j+1}/x_j\to\infty$ so that the integral over $[x_j,x_{j+1})$
is at least $j^\tau$. Then $h(t)=j$ on these intervals is slowly varying,
$h\uparrow\infty$, and $L=\ell h$ still satisfies the integral divergence condition.
Define $L$ to be positive on the initial interval as well. For $A>1$, let $V(A)$ be
the symmetric four-point noise law
\[
 \PP\{w=\pm A\}=A^{-\beta}/2,\qquad
 \PP\{w=\pm s_A\}=(1-A^{-\beta})/2,\qquad
 s_A^2=\frac{1-A^{2-\beta}}{1-A^{-\beta}}\le1.
\]
It has mean zero, variance one and $\beta$th moment at most two.

Use $\tau=\beta/2$ above. At the beginning of the $m$th pair of control steps, set $t=2m$, $T=t+2$.
Inductively, $y_t\le1$, the zero-loss candidates are exactly those
with $\theta_1\le a_m$, and all remaining relative losses are positive;
initially $a_0=1/4$. Choose the smallest integer power of four $A_m$ with
\[
 A_m\ge\max\{4/a_m,\,2a_m^{-1}T^{1/\beta}\sqrt{L(T)}\},
\]
and use noise with law $V(A_m)$ in the first step and Rademacher noise
in the second, using a new draw at each step.
All $h_\theta(y_t)$ vanish, so the output at the first step equals the noise realization and
$q_{t+1}=q_t$. Both selected estimates are therefore $a_me_1$.
If $y_{t+1}=A_m$, every nonzero
$D_{t+1}(\theta)=(\theta_1A_m-1)_+$ is at least three. Hence
\[
 q_{t+2}(\theta)-q_{t+1}(\theta)
 =D_{t+1}(\theta)\bigl(D_{t+1}(\theta)-2w_{t+2}\bigr)>0
 \quad\text{for }\theta_1>A_m^{-1}.
\]
All other loss increments vanish, yielding $a_{m+1}=A_m^{-1}$ and output
$-\Delta\pm1\le-2$, where $\Delta=a_mA_m-1\ge3$.
Otherwise the next output is $\pm1$ and $a_{m+1}=a_m$.
This completes the induction and verifies that every selected estimate
minimizes the full cumulative least-squares criterion. The positive loss
increments occur only after the corresponding estimate has been used
for control.

While $a_m=a$ stays fixed, success has probability
$p_m=A_m^{-\beta}/2\ge ca^\beta/[T L(T)^{\beta/2}]$ eventually;
rounding increases the threshold by a factor of less than four.
Its sum diverges, and $\prod_m(1-p_m)\le\exp(-\sum_mp_m)=0$.
Consequently successes occur infinitely often almost surely, and
\[
 \Reg_T\ge(a_mA_m-1)^2\ge T^{2/\beta}L(T)
 =T^{2/\beta}\ell(T)h(T)
\]
at their endpoints. This proves~\eqref{eq:closed-sharp-heavy}.

\textit{A common construction for (ii) and (iii).}
Let $\Theta=\{0\}\cup[1,2]^d$ and fix a measurable minimizing selector
of $q_t$. For nonnegative uniformly Lipschitz $c_\nu$ with $c_\nu(0)=0$
and Borel $0\le\varphi_\nu(y)\le C(1+|y|)$, set
\[
 f(\theta,y,u,v)=u+\sin y+
 \sum_{\nu\ge1}\mathbf1_{\{v=2^{-\nu}\}}c_\nu(\theta)\varphi_\nu(y),
 \qquad v=u_{t-1}.
\]
Only one summand can be nonzero, so the model is jointly Borel,
parameter-Lipschitz and uniquely invertible with linear growth.
The preceding bound for the true plant verifies the minimum-phase condition.
Starting with $u_{-1}=y_0=0$, implement each observation through two
control steps:
from $u_{t-1}\le0$, references $\sin y_t+2^{-\nu}$ and $\sin y_{t+1}$ give
\[
 u_t=2^{-\nu},\quad D_t\equiv0;\qquad
 u_{t+1}=-c_\nu(\hat\theta_{t+1})\varphi_\nu(y_{t+1})\le0,
 \quad D_{t+1}(\theta)=c_\nu(\theta)\varphi_\nu(y_{t+1}).
\]
The first step leaves the cumulative loss and its fixed minimizing selector
unchanged, so the label can be selected causally. The reference signals
are bounded in absolute value by two;
neutral observations use two steps with $r=\sin y$, $u=0$.

For small dyadic $a$, partition $[1,2]^d$ into $N=(16a)^{-d}$ cubes $C_i$.
In each cube choose $K\asymp(a/v)^d$ points $z_{i,l}$, $4v$-separated
and within $a/2$ of its center, where $v\ll a$. Set
\[
 M_i(\theta)=(a-\operatorname{dist}_\infty(\theta,C_i))_+,
 \qquad H_l(\theta)=\sum_i(v-\|\theta-z_{i,l}\|_\infty)_+.
\]
Extend these functions by zero at $0$. Their Lipschitz constants are
uniform, $H_l(z_{i,l})=v$, and $M_j(z_{i,l})=a\mathbf1_{\{i=j\}}$.
The supports of the auxiliary functions are disjoint, and at most $2^d$
main-function supports overlap. Assign distinct fixed labels to the functions of every stage.

The auxiliary phase is designed to obtain $q(\theta_i)\le-B$ at one grid point
per cube. Allocate $J=\lceil NB/a^2\rceil$ subsequent main observations,
where $B\ge1$ and $n\asymp NB/a^2$. While $\min q<-B/2$, observe
$M_i$ for the first cube containing the selected minimizer, using
$\varphi_\nu=1$; use neutral observations for the remaining allocated observations.
Every active prediction has $\Delta=a$.

We next establish elementary bounds for this construction. During the
main phase, the noise is Gaussian or Rademacher. It may be generated from independent rows, one per cube,
revealing the next unused entry whenever that cube is selected.
Each noise entry used in chronological order is independent of the past
and has the prescribed distribution.
For a row sum $S_{i,l}$,
$\PP\{|S_{i,l}|>x\}\le2e^{-x^2/(2l)}$.
A union bound over $i\le N$ and $1\le l\le J$ gives
$S_{i,l}^2/l\le C\log n$, except with probability $Cn^{-3}$.

Write $A=\sup_\theta|q_{\mathrm{start}}(\theta)|$ and suppose the
auxiliary loss is uniformly bounded by $C(B+\log n)$, as verified below.
Each main term is at least $-S_{i,l}^2/l$, with zero-count terms omitted.
At a selected minimizer, $M_i=a$ and $q\le0$, so the bounded number of overlapping supports gives
$la^2-2aS_{i,l}\le C(A+B+\log n)$. Young's inequality then bounds
$la^2$ by the same quantity, also after the last visit.
Bounding all main terms by the same quantity yields
\begin{equation}\label{eq:sharp-history-recursion}
 \sup_\theta|q_{\mathrm{end}}(\theta)|\le C(A+B+\log n).
\end{equation}
This estimate holds regardless of whether the auxiliary phase succeeds.

If the auxiliary phase succeeds and the main-phase stopping condition
is reached after $j$ active observations, then $NB/2\le ja^2-2aS_j$, where $S_j$ is the sum of the first
$j$ main-phase noise realizations. Each observation changes the loss of exactly one fixed representative
parameter. For $ja^2<NB/4$, this requires
$S_j<-NB/(8a)$. Conditional on such auxiliary observations, the scalar tail bound
and a union bound give
\[
 \PP\{\Reg_{\mathrm{stage}}<NB/4\mid\text{auxiliary history}\}
 \le 2J e^{-NB/32}\le Cn e^{-cNB}.
\]
If the stopping condition is never reached, all $J$ observations are active and the
regret is at least $NB$.

\textit{(ii) Gaussian noise.}
Use independent standard Gaussian noise throughout and $\varphi_\nu=1$.
Take
\[
 v=a^{2(d+1)/(d+2)}/16,\qquad
 m=\lceil v^{-2}\log K/100\rceil,\qquad B=mv^2,\qquad n=Km,
\]
and observe each $H_l$ $m$ times. At every $z_{i,l}$ the auxiliary
loss is $B-2\sqrt B Z_l$, with independent standard normal $Z_l$.
The elementary identity
$\PP\{\max_l Z_l<x\}=(1-\PP\{Z_1\ge x\})^K$
and Gaussian tails give a minimum at most $-8B$, except with probability
$e^{-K^c}$. Also $\max_l|Z_l|\le C\sqrt{\log n}$ except with
probability $Cn^{-3}$; since each parameter activates at most one
$H_l$, this gives the required uniform auxiliary loss bound.
Substituting these choices gives
\begin{equation}\label{eq:sharp-gaussian-scaling}
 B\asymp\log(1/a),\qquad n\asymp a^{-(d+2)}\log(1/a),\qquad
 NB\asymp n^{d/(d+2)}(\log n)^{2/(d+2)}.
\end{equation}
Choose $a_k=2^{-2^{k^2}}$, starting sufficiently late. All the preceding
failure probabilities are summable. Borel--Cantelli and
\eqref{eq:sharp-history-recursion} show that the loss accumulated before the stage is $o(B_k)$,
so every sufficiently late stage has regret at least $cN_kB_k$.
Its endpoint satisfies $T_k\asymp n_k$, proving the positive limsup
in~\eqref{eq:closed-sharp-gaussian}. Theorem~\ref{thm:rates} gives the upper bound.

\textit{(iii) Intermediate finite moments.}
Choose $L=\ell h$ using the construction in part~\textup{(i)}, now with $\tau=\tau_e$.
Starting sufficiently late, take $n_k=\lceil\rho^k\rceil$ and dyadic
$a_k^2\asymp n_k^{-\kappa_e}L(n_k)$. Set
\[
 B=na^{d+2},\qquad v=(a^dB^{\beta/2}/n)^{1/(\beta+d)},\qquad
 A_*=\sqrt B/v,\qquad m=\lceil A_*^\beta\rceil.
\]
The strict inequalities for this regime give $v/a\to0$. Moreover,
$B\asymp n^{1-(d+2)\kappa_e/2}L(n)^{(d+2)/2}\gg\log n$;
also $Km\asymp NB/a^2\asymp n$.

For auxiliary labels use $\varphi(y)=A_*\mathbf1_{\{||y|-A_*|\le3\}}$,
which satisfies a uniform linear-growth bound. Allocate $m$ observations to each
$H_l$. Use preparation noise $V(A_*)$ until its first large value;
then use measurement noise $V(5\sqrt B)$ and fill the remaining
allocated observations with neutral ones. All other noise is Rademacher.
Only large preparations activate $\varphi$, because the output
at the preparation step differs from the noise realization by at most two. Each parameter has
at most one nonzero auxiliary measurement, of magnitude at most $\sqrt B$;
hence the auxiliary loss is deterministically bounded by $CB$.
A positive large measurement gives loss $-9B$ at every $z_{i,l}$.

Generate the auxiliary draws independently for different labels and stages.
In each label the probability of a large preparation is
$1-(1-A_*^{-\beta})^m\ge1-e^{-1}$.
Its probability $\pi_k$ of a positive large measurement therefore satisfies
$\pi_k\asymp B^{-\beta/2}$. The corresponding auxiliary success events
$E_k$ are independent, with
\[
 \PP(E_k)=1-(1-\pi_k)^K
 \asymp KB^{-\beta/2}
 \asymp(n^{\kappa_e}a^2)^{-\tau_e}
 \asymp L(n_k)^{-\tau_e}\longrightarrow0.
\]
The middle equivalence follows by substituting $B=na^{d+2}$ and $v$;
the first uses $KB^{-\beta/2}\to0$.
The divergent integral implies $\sum_k\PP(E_k)=\infty$, so the
second Borel--Cantelli lemma implies that $E_k$ occurs infinitely often
almost surely.

Choose $\rho$ so large that $B_{k+1}\ge4CB_k$ eventually, with $C$
from~\eqref{eq:sharp-history-recursion}. Its failures are summable;
writing $A_k=\sup_\theta|q_{\mathrm{end},k}(\theta)|$, the recursion gives
$A_{k-1}\le B_k$ eventually. On every subsequent $E_k$, the decrease in auxiliary loss therefore
exceeds the loss accumulated before the stage. The main-phase failure bounds
$Cn_ke^{-cN_kB_k}$ are also summable, so infinitely many stages incur
regret at least $cN_kB_k$. The total number of steps allocated to each stage is $\asymp n_k$,
giving $T_k\asymp n_k$ and
\[
 \frac{\Reg_{T_k}}{T_k^{1-\kappa_e}\ell(T_k)}
 \ge c\frac{L(n_k)}{\ell(n_k)}=ch(n_k)\longrightarrow\infty.
\]
All coefficients and labels are fixed in advance. The reference signals
and the admissible conditional noise laws determine the resulting
closed-loop observations.
\end{proof}

\section{Proof of Lemmas \ref{lem:proof-noise} and~\ref{lem:offset}}\label{app:offset-proof}
\subsection{Proof of Lemma \ref{lem:proof-noise}}
\begin{proof}
The martingale strong law gives $W_T/T\to\sigma^2$, and Markov's inequality together with the Borel--Cantelli lemma gives
the noise-peak estimate in~\eqref{eq:noiselln}.
By Theorem 1.3.10 in \cite{guo2020}, $N_T=o(A_T)$ on $\{A_T\to\infty\}$; the predictable cutoffs
$\mathbf1_{\{A_{t+1}\le j\}}$ give convergence when $A_\infty<\infty$.
Thus $|N_T|\le A_T+K(\omega)$ after increasing $K(\omega)$ to cover
the finite initial segment. Expanding the square proves
\eqref{eq:noise-tracking-transfer}. For $2<\beta\le4$, applying Theorem 2.7 and Lemma 2.4 in \cite{chenguo1991} gives
the first bound in~\eqref{eq:noise-rates}.
For $\beta>4$, apply Theorem~1.3.10(ii) in~\cite{guo2020}
to the martingale differences $w_{t+1}^2-\sigma^2$ with unit
coefficients; their conditional $\beta/2$th moments are uniformly
finite almost surely. This gives the second bound in~\eqref{eq:noise-rates}.
Dyadic interpolation and moment localization extend the first bound
to all $T$.

\end{proof}
\subsection{Proof of Lemma \ref{lem:offset}}
\begin{proof}
\textit{Part (i):}
First stop the noise predictably before its conditional
$\beta$th moment exceeds a fixed integer. Markov's inequality and
Borel--Cantelli, applied at every positive rational multiple of
\(
 \varphi(n)=(n+2)^{1/\beta}\sqrt{\log(n+2)},
\)
give $w_n=o(\varphi(n))$ almost surely, since $\beta/2>1$.
This envelope satisfies the growth condition in
\cite[Lemma~3.1]{huangguo1990doublearray}.

For each dyadic $L\ge4$, fix a deterministic $L^{-2}$-net
$\mathcal G_L$ of $\Theta$. For $\theta\in\mathcal G_L$, define
\[
 \begin{aligned}
 d_s^{L,\theta}&=D_s(\theta)
                  \mathbf1_{\{\mathcal E_{s+1}\le L\}},
 & A_t^{L,\theta}&=\sum_{s<t}(d_s^{L,\theta})^2,\\
 x_s^{L,\theta}&=\frac{d_s^{L,\theta}}
                         {\sqrt{1+A_{s+1}^{L,\theta}}}.
 \end{aligned}
\]
These coefficients are predictable, $|x_s^{L,\theta}|\le1$, and
\(
 \sum_{s<N}(x_s^{L,\theta})^2
 \le\log(1+A_N^{L,\theta})\le C\log N,
 L\le N.
\)
Enumerate the pairs $(L,\theta)$ in increasing dyadic $L$.
The pairs with $L\le N$ form a deterministic prefix of
$O(N^{2d})$ rows. The polynomial-array extension in
\cite{huangguo1990doublearray}
therefore gives
\[
 G_N\triangleq
 \max_{\substack{L\le N,\ L\text{ dyadic}\\\theta\in\mathcal G_L}}
 \max_{t\le N}\left|\sum_{s<t}x_s^{L,\theta}w_{s+1}\right|
 =O(\log N)+o(\varphi(N)\log N)
 \quad\text{a.s.}
\]
Abel summation, using the increasing normalization factor
$\sqrt{1+A_t^{L,\theta}}$, yields
\[
 \left|\sum_{s<t}d_s^{L,\theta}w_{s+1}\right|
 \le2\sqrt{1+A_t^{L,\theta}}\,G_N.
\]
For dyadic $N$, take $L=N$ and $\mathcal E_t\le N$; the stopping rule
has then removed no term before $t$, and $t\le N$. Hence
\[
 \max_{\theta\in\mathcal G_N}
 \{2M_t(\theta)-\frac34S_t(\theta)\}
 \le C(1+G_N^2)=O(\Phi(N)).
\]
The parameter Lipschitz bound and $W_N=O(N)$ imply that replacing
any $\theta\in\Theta$ by its nearest point in $\mathcal G_N$
changes $M_t$ and $S_t$ by $O(N^{-1})$, uniformly over
$\mathcal E_t\le N$. Choosing the dyadic
$\mathcal E_t\le N<2\mathcal E_t$ and removing the
conditional-moment cutoff over the positive integers prove
\eqref{eq:prediction-offset}. The same noise envelope gives
$\max_{s\le T}w_s^2=O(T^{2/\beta}\log(T+2))$; the least-squares comparison
then gives~\eqref{eq:prediction-history-coarse}.

\textit{Part (ii):}
\textit{Local decomposition and residual moment bounds.}
Let $c_t$ be the point nearest zero in the interval
\(
 [\min_{\theta\in\mathcal C}D_t(\theta),
  \max_{\theta\in\mathcal C}D_t(\theta)].
\)
Set $v_t(\theta)=D_t(\theta)-c_t$. Compactness and continuity
ensure measurability, and
\[
 c_tv_t(\theta)\ge0,\qquad |v_t(\theta)|\le L_fa\zeta_t,
 \qquad D_t(\theta)^2\ge c_t^2+v_t(\theta)^2.
\]
Thus the offset is bounded by the sum of the offsets associated with
the scalar sequence $c_t$ and the residual sequence $v_t$. For the former, under the localized moment bound, Theorems 1.3.6 and 1.3.7 in \cite{guo2020}
give a maximal $b$th-moment bound $CL^{b/2}$ when stopping before
the coefficient energy, including the current coefficient, exceeds $L$.
Applying Markov's inequality on the energy ranges $[0,H]$ and
$(2^jH,2^{j+1}H]$, and summing over $j\ge0$, yields
\begin{equation}\label{eq:scalar-offset-tail}
 \PP\!\left\{\max_{t\le N}
 \left(2\sum_{s<t}c_sw_{s+1}-\frac12\sum_{s<t}c_s^2\right)>H\right\}
 \le CH^{-b/2},\qquad H\ge1.
\end{equation}
Since $b>2$, the integrated tail is bounded by $CH^{1-b/2}$.

For the residual offset, define
\[
 q_t(H)=\sup\left\{|v_t(\theta)|:\theta\in\mathcal C,
                 \sum_{s\le t}v_s(\theta)^2\le2H\right\},
 \qquad \sup\varnothing=0.
\]
The current coefficient is included, so this envelope is predictable
and $q_t(H)\le\min\{L_fa\zeta_t,\sqrt{2H}\}$.
Cover $\mathcal C$ at resolution $u/(2L_f)$. At times with
$q_t(H)>u\zeta_t$, a maximizing parameter chosen at a cell's last
visit retains at least half the earlier residual magnitudes.
The corresponding energy includes the contribution from that last visit. Consequently
\[
 \sum_{\{t:q_t>u\zeta_t\}}q_t^2\le CH(a/u)^d,\qquad
 \sum_{\{t:q_t>u\zeta_t\}}\zeta_t^2
 \le\min\{N,CHa^du^{-d-2}\}.
\]
Split the sum at $q_t=u\zeta_t$ and balance the two resulting bounds, using
$u^{m+d}=a^dH^{m/2}/\mathcal E_{m,N}$. Together with the direct
envelope bound this gives
\begin{equation}\label{eq:unified-active-moment}
 \sum_{t<N}q_t(H)^b\le C\min\left\{
 a^m\mathcal E_{m,N}H^{(b-m)/2},
 a^{md/(m+d)}\mathcal E_{m,N}^{d/(m+d)}
 H^{b/2-md/[2(m+d)]}\right\},\quad 2\le m\le b.
\end{equation}
Alternatively, integrating the weighted counting bound, with
$\zeta_t^{b-2}\le P^{(b-2)/2}$, gives
\begin{equation}\label{eq:unified-peak-budget}
 \sum_{t<N}q_t(H)^b\le CP^{(b-2)/2}
 \min\{Na^b,
 a^{bd/(d+2)}N^{1-b/(d+2)}H^{b/(d+2)}\},\quad 2<b<d+2.
\end{equation}

\textit{A uniform estimate based on common truncation.}
The following estimate is also used for logarithmic thresholds.
Fix $s,B\ge1$ and a predictable $\bar q_t\ge q_t(4s)$, and write
\(
 z_t=\bar q_t|w_{t+1}|
           \mathbf1_{\{\bar q_t|w_{t+1}|>B\}}.
\)
If $\sum_{t<N}\Et z_t\le s$, as may be enforced by a further
predictable stop, then, for $2\le b\le\beta$ and fixed $D>1$,
\begin{equation}\label{eq:offset-truncation}
 \begin{aligned}
 &\PP\!\left\{
 \sup_{\substack{t\le N,\,\theta\in\mathcal C\\
                    \sum_{i<t}v_i(\theta)^2\le s}}
 \left|\sum_{i<t}v_i(\theta)w_{i+1}\right|>Cs,
 \ W_N\le N^D\right\}\\
 &\quad\le CN^C\exp\!\left\{-\frac{cs\log(1+B)}B\right\}
 +\frac C{s^b}\left\{
 \E\left(\sum_{t<N}\Et z_t^2\right)^{b/2}
 +\sum_{t<N}\E z_t^b\right\}.
 \end{aligned}
\end{equation}
To obtain this bound, truncate all products at the common threshold
$\bar q_t|w_{t+1}|\le B$ and subtract their conditional means. For each deterministic candidate, stop before its residual
energy exceeds $4s$. The retained martingale has increments bounded
by $2B$ and variance at most $C\sigma^2s$, so the
Bennett--Freedman exponential-supermartingale inequality gives
the displayed exponential term. Related uses of exponential martingale
estimates appear in the offset analyses
in~\cite{stamouli2024pem,liguo-offonline}.
The discarded absolute products and their centering are bounded by
$\sum z_t+\sum\Et z_t$. The maximal martingale moment inequality
for $z_t-\Et z_t$ gives the remaining term.

To obtain a bound uniform over the parameter region, use a deterministic $N^{-K}$-net before imposing
candidate-dependent stops. On $W_N\le N^D$, the original
martingales and square roots of their energies change by at most
$CN^{-K}\sqrt{NW_N}$ and $CN^{-K}\sqrt N$, respectively.
Choose $K>(D+1)/2+1$. A candidate of energy at most $s$ then has
a neighboring net point of energy at most $2s$, whose stop at $4s$
has not been triggered. A union bound accounts for $N^C$ in
\eqref{eq:offset-truncation}.

Alternatively, take $\bar q_t=q_t(4s)$ and exclude the event on which
at least one product is truncated. Its probability is at most
$CB^{-b}\E\sum q_t(4s)^b$. If $p_t$ is the conditional truncation
probability, the centering correction is at most
$C\sqrt{s\sum p_t}$; the probability that it exceeds the required threshold is at most
$CB^{-b}s^{-1}\E\sum q_t(4s)^b$.
Thus, without the conditional-mean restriction, the moment term
in~\eqref{eq:offset-truncation} may instead be replaced by
$CB^{-b}\E\sum q_t(4s)^b$.

\textit{Polynomial thresholds.}
For $H\ge N^\eta$, use the preceding alternative bound with
$B=\max\{1,s/K_0\}$ on dyadic residual-energy bands
$s\ge x\ge H$. Insert~\eqref{eq:unified-active-moment}
or~\eqref{eq:unified-peak-budget}, sum the geometric bounds,
and integrate in $x$. Equation~\eqref{eq:scalar-offset-tail}
contributes $CH^{1-b/2}$. Since $U_{\mathcal C}\le2W_N$,
the contribution from the excluded event is bounded by
\[
 \E[U_{\mathcal C}\mathbf1_{\{W_N>N^D\}}]
 \le CN^{b/2+D(1-b/2)}.
\]
Choose $D$ first, then the net resolution and $K_0$, so that this contribution
and the integrated exponential term are at most $N^{-A_0}$.
This proves~\eqref{eq:regional-polynomial}--\eqref{eq:regional-peak}.

\end{proof}

\bibliographystyle{IEEEtran}
\bibliography{NLS}

\end{document}